\documentclass[10pt]{bmc_article}

\usepackage{cite} 
\usepackage{url}  
\usepackage{ifthen}  
\usepackage{multicol}   
\usepackage[utf8]{inputenc} 
\usepackage{amssymb}
\usepackage{graphicx}
\usepackage{color}

\newboolean{publ}

\newenvironment{bmcformat}{\fussy\setboolean{publ}{true}}{\fussy}

\newcommand\cmt{\mbox{\boldmath$M$}}
\newcommand\vecBold[1]{\mbox{\boldmath$#1$}}
\newcommand{\mr}{\mathbb{R}} 
\newcommand\lift{l}
\newcommand\cp{p}
\newcommand\rp{r}

\begin{document}
\begin{bmcformat}



\title{Predicted selective increase of cortical magnification due to cortical folding}


\author{Markus A Dahlem\correspondingauthor$^{1,2}$%
       \email{Markus A Dahlem\correspondingauthor - dahlem@physik.tu-berlin.de}%
      and
         Jan Tusch$^3$%
         \email{Jan Tusch - tusch@isg.cs.uni-magdeburg.de}
      }


\address{%
    \iid(1) Institut f{\"u}r Physik, Humboldt Universit{\"a}t zu Berlin, Germany\\
    \iid(2) Institut f{\"u}r Theoretische Physik, Technische Universit{\"a}t Berlin, Germany\\
    \iid(3) Department of Simulation and Graphics
Faculty of Computer Science, University of Magdeburg, Germany 
}%

\maketitle


\begin{abstract} 
The cortical magnification matrix  \vecBold{M} is introduced founded on a notion
similar to that of the scalar cortical magnification factor $M$.  Unlike $M$,
this matrix is suitable to describe anisotropy in cortical magnification, which
is of particular interest in the highly gyrified human cerebral cortex.  The
advantage of our tensor method over other surface-based 3D methods to explore
cortical morphometry is that \vecBold{M} expresses cortical quantities in the
corresponding sensory space.  It allows us to investigate the spatial relation
between sensory function and anatomical structure.  To this end, we consider
the calcarine sulcus (CS) as an anatomical landmark for the primary visual
cortex (V1). We found that a stereotypically formed 3D model of V1 compared to
a flat model explains an excess of cortical tissue for the representation of
visual information coming from the horizon of the visual field. This suggests
that the intrinsic geometry of this sulcus is adapted to encephalize a
particular function along the horizon.  Since visual functions are assumed to
be $M$-scaled, cortical folding can serve as an anatomical basis for increased
functionality  on the horizon similar to a retinal specialization known as
visual streak, which is found in animals with lower encephalization. Thus, the
gain of surface area by cortical folding links anatomical structure to cortical
function in a previously unrecognized way, which may guide sulci development.
\end{abstract}

\ifthenelse{\boolean{publ}}{\begin{multicols}{2}}{}


\section{Introduction}
\label{sec:introduction}

The patterns of the highly folded surface of the cerebral cortex are prominent
features of the human brain (Fig.~1~{\sf A}).  Primarily, folding permits a
larger cerebral cortex surface area to fit inside the skull.  {However, folding
ensures that the additional surface area is not homogeneously distributed, if
the surface becomes intrinsically curved.} The surface gain is spatially
concentrated in certain cortical regions. The cortex represents sensory
information in distinguishable fields and therefore the question of the
relationship between anatomical structure and sensory function is naturally
given.  We utilize methods of continuum mechanics and complex analysis to
explore this relationship in the cortex.

Many  studies of human cortical architecture show that sensory and motoric
fields have some relationship to the gross sulcal and gyral morphology,
although a substantial variability in both the size and location is observed
\cite{Rajkowska1995,Thompson1996,Roland1998,Amunts1999}. In a few cases very
precise correlations between sulci and functional entities could be
demonstrated. Motor cortex can be identified by the position of the central
sulcus \cite{Lotze2000} and the primary auditory cortex has a clear spatial
relationship with Heschl's gyrus \cite{Gaschler-Markefski1997,Rademacher2001}.
The most reliable relation is, however, the calcarine sulcus (CS) as a landmark
of the primary visual cortex (V1) \cite{Stensaas1974,Gilissen1996,Andrews1997}.
Anatomical identification is also quite reliable for visual areas outside V1,
e.\,g., V5 lies at the intersection of the ascending limb of the inferior
temporal sulcus and the lateral occipital sulcus \cite{Watson1993,Walters2003}. 

In this paper, we consider V1 because not only its structural anatomy but also
its functional retinotopy, i.e., the spatial organization of the neuronal
responses to visual stimuli (see below), is well studied in this field
\cite{Schneider1993,Engel1994,Sereno1995,DeYoe1996,Engel1997,Qiu2006}.  The CS,
where V1 is located,  begins near the occipital pole on the medial surface of a
hemisphere. It continues towards the posterior end of the corpus callosum
(Fig.~1 {\sf A}).  We will present evidence that the 3D form of the CS
indicates a selective magnification of the horizon of the visual field that is
neither accounted for in  standard  retinotopic maps nor reflected in the
density of retinal ganglion cells, { although a modest increase of retinal cell
density, a so called visual streak, can be found
\cite{Stone1981,Wassle1989,Curcio1990}}.

Already in 1984 Rovamo and Virsu \cite{Rovamo1984} have noted that an locally
isotropic (independent of the direction of measurement) cortical magnification
that is also symmetric with respect to the meridians can be better approximated
by taking the unfolded convex 3D form of the cortex into account. This does not
imply that cortical magnification has to be strictly locally isotropic but
curvature affects the overall layout.  Recently, the influence of cortical
folding in primate did also take into account the concave folds
\cite{Rajimehr2009}. We apply tensor analysis to investigate how these
symmetries, i.e., isotropy and meridional symmetry, relate to the gross folding
pattern, in particular the concave fold of CS that creates additional cortical
space for the representation of the visual field close to the horizon.  To this
end, we compare intrinsically flat and  curved surfaces of V1 and investigate
how the 3D form affects cortical magnification.  We  propose that in particular
a horizontal stripe gains additional cortical space.  The shape and location of
this stripe is similar  to the visual streak as a retinal specialization found
to be very pronounced in some animals
\cite{Steinberg1973,Picanco-Diniz1991,Peichl1992,Vidyasagar1992,Guo2000,Arrese2003,Calderone2003}.
This suggests a link between cortical folding, $M$-scaling and the functional
development of cerebral sulci.  

First evidence for the predicted selective cortical magnification of the visual
horizon can be found in the literature: perceptual filling-in
\cite{Sakaguchi2003} and travelling migraine scotoma \cite{Lashley1941}. The
big advantage of psychophysical methods that measure scotoma is that neither
method requires surface reconstruction to measure cortical magnification.  But
it is difficult to get reliable quantitative data, because these methods
involve psychophysical investigations with subjective evaluations from probands
and patients.  {Furthermore, there is a brief report of such a phenomenon
\cite{Janik2003}}. Only computationally-intensive methods, which allow precise
surface-based morphometry using anatomical magnetic resonance imaging and
functional magnetic resonance imaging (MRI and fMRI)
\cite{ChungEvans2003,Qiu2006,Henriksson2012}, can provide a direct test of our
predicted correlation between CS, as an anatomical cortical landmark, and
increased $M$, as a functional cortical measure.

\begin{figure*}
\centerline{\includegraphics[width=0.75\textwidth]{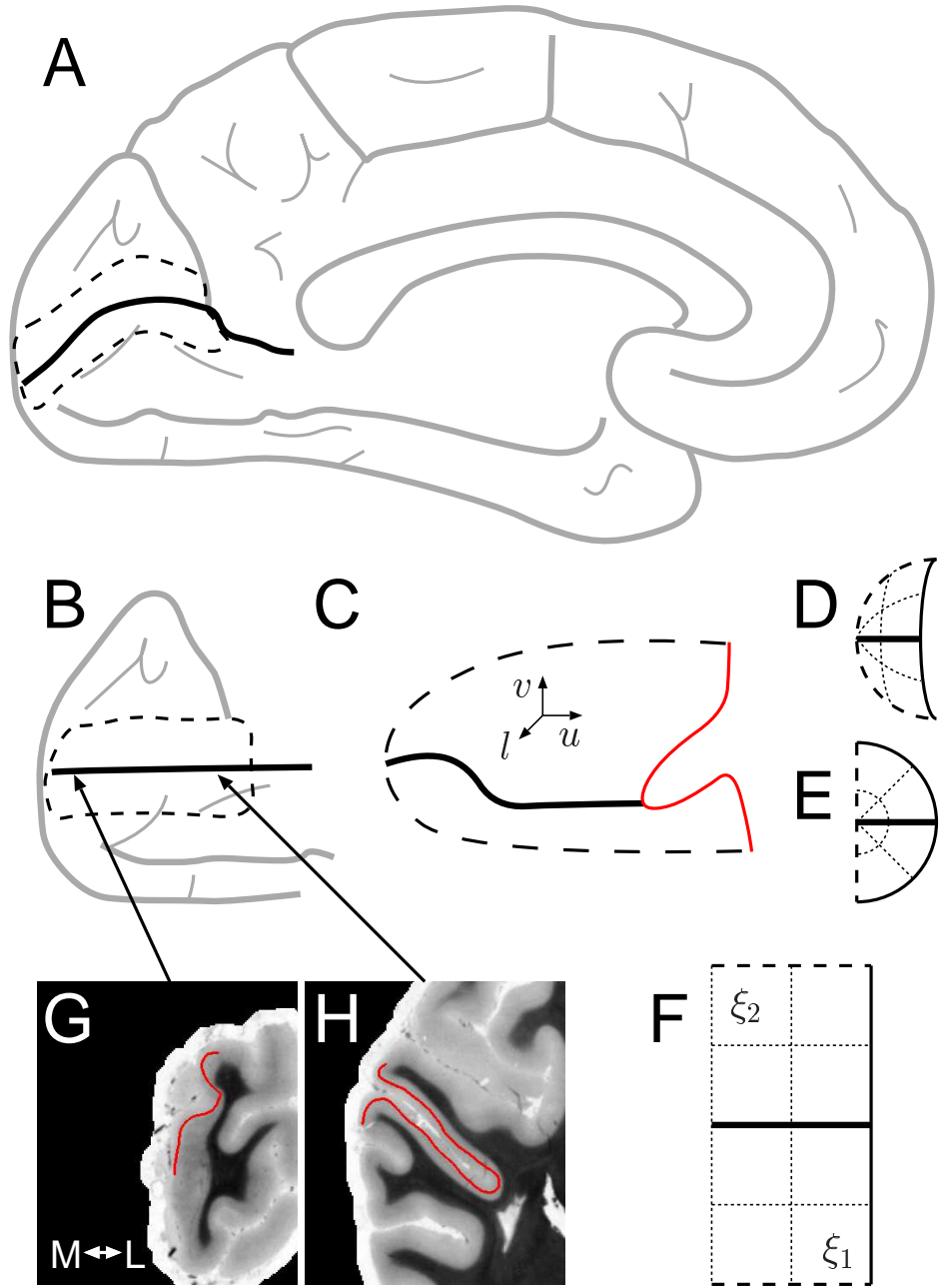}}
\caption{ 
({\sf A}) The calcarine sulcus (CS) begins near the occipital pole on the
medial surface of a cortical hemisphere and continues toward the posterior end
of the corpus callosum (thick black line). This is the site of V1. Its border
is marked with a dashed line.  ({\sf B}) Medial side of the occipital pole, as
in ({\sf A}) but with straight course of CS. ({\sf C}) The approximate 3D form
of V1 in the $(u,v,l)$ coordinates (see text). The representations of the two
vertical meridians are dashed and the fundus of CS is marked with a thick black
line.  ({\sf D}) The retinal surface is approximated by a quarter of a sphere,
or ({\sf E}) by a flat disc, as assumed in many experimental data.  ({\sf F}) A
Cartesian reference coordinate system of the visual field (${\xi_1}, {\xi_2}$).
({\sf G}) and ({\sf H}) Coronal section through V1. The red line traces the
stria of Gennari. The {\sf M-L} axis indicates the medial to lateral axis.
\label{fig:medialStraightCSReferenceFrame}
}
\end{figure*}

\section{Material and Methods}
\label{sec:materialAndMethods}

We use concepts from continuum mechanics  to describe the retino-cortical
map\cite{Salencon}.  In the Lagrangian description of continuum mechanics, the
map is seen as a deformation rather than a transformation.  This description
uses some coordinate system in the sensory space  as reference configuration to
express every quantity in the deformed configuration, i.\,e.,  as the cortical
map of the sensory space.  The deformation language is intuitive but needs to
be adapted for our purposes. We emphasize that we take  mostly a kinematic
approach. No attention is paid to the dynamics of the creation of a retinotopic
map { other than that we compare two discrete stages of V1, one which is
intrinsically flat and one that evolved from the latter such that it is
intrinsically curved with a major sulcus}. The Lagrangian description is also
suited for a dynamical description  continuous in time of map development, for
example with self-organizing neural networks \cite{tusch04}.

Let us start by briefly clarifying the terminology, in particular related to
cortical mapping and curvature; linear and areal cortical magnification; the
naming of the visual coordinate system; and, finally, tensor versus matrix
methods.

Conformal maps  allow us to describe cortical magnification by a scalar field,
simply referred to as cortical magnification factor.  In fact, we start to
consider conformal maps between domains in the complex plane, that is, we
consider intrinsically flat domains. Conformal maps also exist between
intrinsically curved domains.  Note that throughout the manuscript, we do not
consider extrinsic curvature unless we say so. Of course, to fit larger
cerebral cortex surface area inside the skull, nature can make primarily use of
extrinsic curvature, but not exclusively (unless the cortex would be a
cylindrically scrolled structure). In the following, we will refer to V1  as
being curved (''curved V1''), if it is a surface with intrinsic curvature at
least at some locations within V1, i.e., locations that have a non vanishing
Gaussian curvature. As a consequence, a curved V1 must be embedded in 3D even
if it is mostly flat.  And we will refer to V1  as a ''flat V1'', if it is an
intrinsically flat surface, i.e., the Gaussian curvature is zero everywhere.

It can be important to distinguish linear and areal cortical magnification, if
we assume $M$-scaling, i.\,e., the fact that a measured quantity remains
qualitatively similar across the entire visual domain when magnified in inverse
proportion to $M$. For example, the time required for a scotoma of certain size
to fade and become replaced by its background (perceptual filling-in, see
discussion) can be proportional either to linear or areal cortical
magnification. The underlying mechanism could involve merely the scotoma
diameter, which would suggest linear $M$-scaling or involve the total area
covered by the scotoma, which would suggest areal $M$-scaling.  The term
$M$-scaling is unfortunately sometimes used without mentioning which is meant.
Note that the distinction between linear or areal cortical magnification is in
cases where $M$ is a scalar field not essential as these factors can easily be
converted, but one still must specify which method was used and in which units
$M$ is given (see next section).

To specify loci in the visual hemifield, a polar coordinate system with its
origin in the viewer's fixation point is used (Fig.~1 {\sf E}).  The coordinate
lines at fixed polar angles are  called meridians, with the horizontal meridian
(HM) as the reference set to be zero degrees (bold solid line in Fig.~1 {\sf
E}). Meridians increase and decrease in the upper and lower visual field
quadrant to 90 degrees going anticlockwise and clockwise, respectively, until
the two vertical meridians (bold dashed lines in Fig.~1 {\sf E}) bound the
visual hemifield.

We will express  the generalized  cortical magnification in terms of a matrix
\vecBold{M}.  We want to stress that cortical magnification is a concept independent
of the retinal and cortical coordinate systems chosen to represent it.
Therefore, \vecBold{M} actually is a  cortical magnification tensor. It may lead to
some confusion, when we later introduce \vecBold{M} based on the cortical metric in
the sensory coordinate system.  We need to consider some sensory coordinate
system, which merely reflects the fact that cortical magnification is not
exclusively  a property of the intrinsic geometry of the cortex but of the
sensory map in the cortex. In mathematical terms, \vecBold{M} is the square root of
the matrix expressing the Riemannian metric of V1 in the sensory polar
coordinates. 

The Lagrangian formulation of a such maps is a tensor approach. As an
alternative to Riemannian geometry, it provides by definition some retinal
parameterization of the cortical surface that is needed for the concept of
magnification from retina to cortex.  In contrast, the cortical metric in an
arbitrary parameterization can be used, for example, to define the
Laplace-Beltrami operator as a generalization of the Laplace operator and much
effort has gone in the development of computationally advantageous
parameterizations in this regard.   However, since the cortical magnification
is usually discussed referring to meridians and eccentricity, it is natural to
use exclusively this polar coordinate system. Therefore, we refer to the
generalized \vecBold{M} as the cortical magnification  matrix, and avoid the
potentially daunting term tensor.

\section{Results}
\label{sec:results}

We will first define the contour of an intrinsically flat V1.  Then we will
define the cortical magnification matrix and use this concept firstly to
compare cortical magnification along different radial directions (from
horizontal to vertical) on the flat V1 and show that the vertical direction has
an increased cortical magnification.  Let us emphasize that retinotopic maps
that are assumed to be intrinsically flat can still be extrinsically curved. We
will define an intrinsically curved V1 starting from an extrinsically flat V1
by plausible deformations.  In a second step, the retinal coordinate system is
defined on this curved V1, preserving some symmetry constraints while relaxing
others. Finally, we use the concept of the cortical magnification matrix to
predicted a selective increase of cortical magnification along the horizon due
to cortical folding.

\begin{figure*}[!tpb]
\centerline{\includegraphics[width=\textwidth]{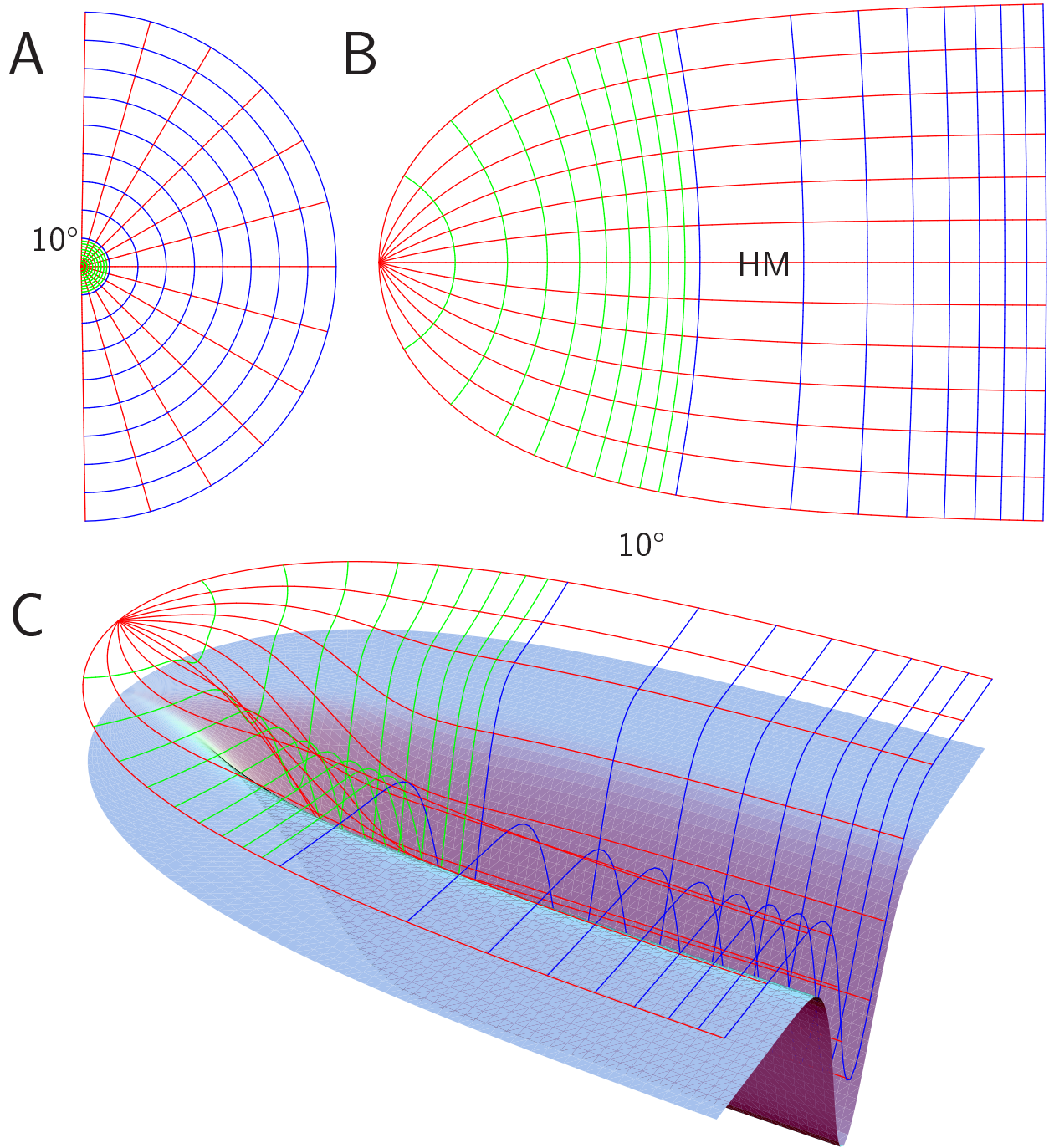}}
\caption{
({\sf A}) Retinal grid which serves as a reference frame in the Lagrangian
formulation of retinotopy. In the inner 9$^\circ$, iso-eccentricity lines are
drawn in 1$^\circ$-steps in green. From 10$^\circ$ to 90$^\circ$
iso-eccentricity lines are drawn in 10$^\circ$-steps in blue. Meridians are
drawn in 15$^\circ$-steps in red. ({\sf B}) The neural representation of the
retinal grid mapped with Eq.\ \ref{eq:logzp1} serves as a flat model of V1.
({\sf C}) The blue surface serves as a curved model of V1. It is created by
forming {\sf B} with lift functions (see Fig.~1 {\sf G} and {\sf H}) in the
third dimension. The neural representation of the retinal grid on this surface
is placed such that meridians are equidistantly spaced.  \label{fig:retinotopy}
}
\end{figure*}

\subsection{Symmetries in the retinotopic map}

As a starting point, we need the contour of the area of V1. The contour of  V1
can be defined by the bounded domain of the visual hemifield that V1 neurally
represents. Simply speaking, V1 has a map of one visual hemifield with a
nonuniform map scale. This scale is called cortical magnification.  This map,
or rather the retinotopic mapping function governs the contour of V1, because
it maps the bounded domain of the visual hemifield, approximated by a half
disc,  that extents to a visual eccentricity of nearly up to 90$^\circ$ (see
Fig.~2A). We will contrast some empirical observations with deductive reasoning
based on plausible symmetry conditions on the retinotopic map to obtain the
contour of  V1 and to introduce these symmetries that guide us through the
following sections.

The retinotopic  map is usually expressed by the cortical magnification factor
$M$.  For a mapping function from an one dimensional domain (a line) to
another, e.g., along a single meridian, it is sufficient knowing the value of
$M$ on this domain. Note, that the term {\it linear} cortical magnification
factor (see previous Section) refers to the one dimensional domain. This can
cause confusion, because $M$ is often modeled as an inverse linear function of
a single coordinate, the retinal eccentricity $\theta$. A fact that will be
important in the following brief analysis of the involved symmetries.

In the one dimensional case $M$ is simply the derivative of the mapping
function.  The mapping function can therefore be determined up to a constant of
integration. This constant can be set to zero, because it describes only a
translational shift.

The complex logarithm provides a standard formulation of retinotopy in a flat
V1.  Fischer \cite{Fischer1973} first suggested this analytic function for the
transformation of the visual field into its neuronal representation. While this
relation was derived from visual inspection of ganglion cell density and
receptive field size distribution assuming these quantities are $M$-scaled, it
is supported  by data of $M(\theta)$ \cite{Daniel1961}, as shown by Schwartz
\cite{Schwartz1977a} with a power law $\frac{1}{b}\theta^{-n}$ fit. The
exponent $n$ being sufficiently close to unity to be replaced by it.  The
meaning of parameter $b$ will be explained below.  Thus the integral equation
reads as \begin{equation} \label{eq:logtheta} u(\theta) = \int M(\theta)
d\theta = \int \frac{1}{b}\theta^{-1}d\theta = \frac{1}{b} \log \theta,
\end{equation} where $u$ is a cortical linear distance, for instance the
distance along the cortical representation of the horizon in the visual field.  

In two dimensions, the situation is more complicated. The exact unity value of
the exponent $n=1$ can even be postulated ab initio from symmetry constraints
for 2D maps. It is therefore important to distinguish these two perspectives.
In 2D, a possible but purely hypothetical function is obtained by generalizing
the real function in Eq.~\ref{eq:logtheta}  to a corresponding complex function
\begin{equation} \label{eq:logz} w = \frac{1}{b} \log z.  \end{equation} The
magnitude of $z$ is the retinal eccentricity $\theta$ and its argument $\phi$
is the azimuth ($z=\theta e^{i\phi}$), and the real and complex parts of $w$
are Cartesian coordinates $(u,v)$, respectively.   To  generalizing the real
function in Eq.\ \ref{eq:logtheta}  to a corresponding complex function implies
two rather obvious constraints on the retinotopic map and one more subtle
constraint.  

Firstly, a complex function implies conformal mapping, that is, cortical
magnification becomes a scalar field. This is a symmetry, namely that cortical
magnification at any location is invariant with respect to direction.
Therefore, cortical magnification is locally isotropic. At least for some
$M$-scaled visual functions this seems to be a natural symmetry requirement,
such as for visual acuity.  Secondly, complex functions are conformal maps
between domains in the complex plane, that is,  intrinsically flat domains,
which, of course, can still be extrinsically curved to accommodate the limited
extend of the skull.  This is another symmetry, namely that the Gaussian
curvature is constant and zero everywhere.

These symmetries, which are given by the constraints of analytical functions,
look rather reasonable. But even if these symmetries were reasonable in the
light of missing or uncertain experimental data (see discussion), they should
not be given preference to further symmetries that could also be found in the
retinotopic map, in particular meridional symmetry of cortical magnification,
i.e., $M$ is invariant unter retinal rotations around the center. 

We will discuss whether meridional symmetry of $M$ should be considered as a
plausible constraint, but let us end this section by emphasizing its severe
consequences.

The real and imaginary parts of an analytic function are conjugate harmonic
functions that solve Laplace's equation. From Laplace's equation in polar
coordinates, it can be easily shown that Eq.~\ref{eq:logz} is harmonic and that
any harmonic function that depends only on the eccentricity $\theta$ must be of
this form.    Alternatively,  this follows from Cauchy–Riemann differential
equations in complex analysis,  which must be satisfied if we assume that the
analytical mapping function is differentiable\cite{Ahlfors}. 

Thus, it is readily shown that any meridional symmetric analytic function must
be of the form of Eq.~\ref{eq:logz}.  Therefore, any retinotopic map that is an
analytic function with a meridional symmetric magnification factor $M$ implies
that cortical magnification is an inverse linear function (in fact, a linear
function with the axis intercept at zero).    So if cortical magnification is
investigated under the following two assumptions: (a) independent on direction
(conformal map), (b) intrinsically flat (with (a) this leads to analytical
functions), (c) meridional symmetry of $M$, one should be aware of the implicit
assumption of inverse linear cortical magnification in the form of Eq.\
\ref{eq:logz}.

Even more importantly, Eq.\ \ref{eq:logz} fails to describe the retino-cortical
map close to the representation of the fovea ($\theta\!<\!1$) because the
foveal point at $\theta\!=\!0$ is a singularity.  To include the foveal region
an offset can be introduced \begin{equation} \label{eq:logzp1} w =
\frac{1}{b}\log (\frac{b}{a} z +1 ).  \end{equation} together with another
parameter $a$.  Eq.\ \ref{eq:logzp1} is not meridional symmetric. It follows
from the reasoning before that a retinotopic map that is based on an analytic
function and includes the  representation of the fixation point ($\theta\!=0$)
cannot be meridional symmetric. 

Both parameters can now be interpreted easily when Eq.\ \ref{eq:logzp1} is
differentiated: $\frac{1}{a}$ is the value of the linear cortical magnification
factor $M$ at the center of the visual field in terms of millimeters of cortex
per degree of visual angle.  The parameter $b$ is the linear growth rate
(slope) of the inverse linear cortical magnification factor on the real axis
(${\phi=0}$), i.\,e., along the horizon.  This map is called monopole map (see
Fig.~2 {\sf A-B}).  Values of the parameters $a$ and $b$ are given in the
literature \cite{Slotnick2001}.  Note that sometimes different parameter are
used, in particular  $w = A\log (z + E_2 )$ with $a=E_2/A$ and $b=1/A$.

We use normalized values in units of $a$. The value of $b$ is  chosen as
0.57265$a$ which would correspond to $a=0.117$ and $b=0.067$ as in
Ref.~\cite{Cowey1974}. When Eq.\ \ref{eq:logz} is used, for example by
\cite{Qiu2006}, there is only parameter $b$. Our dimensionless parameter $a$
can be estimated from the contour  of the flat map.

The inverse linear magnification holds in the map given by Eq.\ \ref{eq:logzp1}
only at the horizon. Due to the shift, none of the other meridians take a
simple course in the complex domain (overlapping with coordinate lines of the
real and imaginary part or the coordinate lines of the absolute value and
argument). Therefore, to investigate the  cortical magnification factor along
other meridians we introduce the cortical magnification matrix, which can
express magnification at  arbitrary points along arbitrary directions most
easily.  Furthermore, unlike the scalar factor $M$, this matrix is also
suitable to describe anisotropy in cortical magnification, which is of
particular interest in the highly gyrified human cerebral cortex.  But it can
also be useful for conformal maps, in which case the matrix at any point can be
transformed to the identity matrix multiplied by a scalar.

\subsection{Generalized cortical magnification}

The natural description for a generalized cortical magnification is the
Lagrangian description.  It uses the retinal configuration as a reference to
express every quantity in the cortical configuration.  Mathematically this is
expressed as \begin{equation} \label{eq:Phi} \Phi:D_R \rightarrow D_{V1}: \rp
\mapsto \cp = \Phi(\rp), \end{equation} where $\rp{}$ and $\cp{}$ are points in
the retinal and cortical domain $D_R$ and  $D_{V1}$, respectively.  The
Jacobian matrix of map $\Phi$ can be interpreted as a deformation gradient
\begin{equation} \label{eq:jacobian} \vecBold{J}_\Phi(\rp)=\nabla \Phi(\rp).
\end{equation} It is a homogeneous transformation tangent to the transformation
$\Phi$ attached at the image of $\rp$.  Caution is needed when the classical
linear cortical magnification factors are derived from the components of
$\vecBold{J}_\Phi$.  Usually non-Cartesian coordinates are utilized to describe
the retinal reference state. Furthermore, $\vecBold{J}_\Phi$ contains a
rotation which carries the retinal directions onto its neural representations.
This rotation introduces problems if cortical magnification  is not isotropic.
To obtain the linear cortical magnification factor in accordance with its
definition \cite{Daniel1961}, one must use the scalar product of a vector in
the considered direction.  The linear cortical magnification factor $M_{v_\rp}$
of a retinal tangent unity vector $v_\rp$ ($\Vert v_\rp \Vert = 1 $) attached
at $\rp$ is the norm of its cortical image  under the homogeneous
transformation: \begin{equation} \label{eq:norm} M_{v_r} = \Vert \nabla
\Phi(v_r)\Vert.  \end{equation} This norm and the underlying scalar product is
defined via the  cortical metric  in retinal parameterization \begin{equation}
\vecBold{g}_{V1}=\vecBold{J}_\Phi^T \vecBold{g}_R \vecBold{J}_\Phi,
\end{equation} where $\vecBold{g}_R$  
$\vecBold{g}_{V1}$ are the metrices of retina and cortex, respectively.  The
cortical metric $\vecBold{g}_{V1}$ leads to the definition of the cortical
magnification matrix.  This matrix is founded on a notion similar to that of
the linear cortical magnification factor $M$, but is much broader in
conception.  Once we have this matrix, we can calculate the linear and cortical
areal magnification factors on the flat and curved model of V1 and compare the
results.  

In continuum mechanics $\vecBold{g}_{V1}$ is called the right Cauchy-Green
tensor and termed $\vecBold{C}$. For the sake of simplicity, we follow Salencon
\cite[p. 44]{Salencon} and adopt the term (cortical) expansion tensor for
$\vecBold{C}$. This name may better convey the meaning of this matrix.  Its
eigenvalues and eigenvectors give the shape and orientation of an ellipsoid
representing an initially spherical infinitesimal area in the retina.
Furthermore this alternative name reminds us that we are not interested in the
cortical metric expressed in an arbitrary coordinate system but in one that is
related to the sensory space. This would be sufficient for other surface-based
3D geometric tasks using, for instance, the Laplace-Beltrami operator to create
flat maps with minimal surface area distortions. Using this particular cortical
surface parameterization, namely some sensory coordinates makes the choice
fundamental. It links the metric to sensory function and it allows us to
describe changes of  the cortical surface from the sensory perspective.

The square root of $\vecBold{C}$ is a pure stretch tensor that is directly
related to the classical concept of cortical magnification and it could
therefore also be termed cortical magnification tensor, yet the term matrix is
probaly more intuitiv in the neuroscience community
\begin{equation} 
\vecBold{M}=\sqrt{\vecBold{C}}.
\end{equation}
The matrix  \vecBold{M} is alternatively obtained from a polar decomposition of
$\vecBold{J}_\Phi \vecBold{J}_\Psi$. The Jacobian matrix $\vecBold{J}_\Psi$ is
the homogeneous transformation tangent to $\Psi$, which maps Cartesian
coordinates to the visual field coordinates, i.\,e., either spherical or polar
(retinal) coordinates.  The distinction between spherical and polar coordinates
should be formally made, since data of the visual field position can be
obtained either with perimetry or campimetry. We can avoid this, however, if we
transform the retinal system into the Cartesian coordinates $({\xi_1},{\xi_2})$
of the visual field (see Fig.\ 1 {\sf G}), because data is often available in
these coordinates, for instance migraine aura scotoma drawn on paper and
subsequently scanned.

\subsection{Linear cortical magnification factor derived from the matrix \vecBold{M}}
\label{sec:lmf}

The linear cortical magnification factor $M_{\vecBold{v}_\rp}$ can be defined
at any point $\rp$ for arbitrary directions $\vecBold{v}$. In experiments,
$M_{\vecBold{v}_\rp}$ is usually measured along a constant meridian, though not
necessarily along this direction.  In conformal maps, e.\,g., the monopole map
(Eq.\ \ref{eq:logzp1}), $M_{\vecBold{v}_\rp}$ depends only on position $\rp$
but not on direction $\vecBold{v}$. 

We compare the $M_{\vecBold{v}_\rp}$ on the  monopole map (flat V1) with the
one on the curved V1 along the $\theta$-direction.  This direction is most
frequently used in the definition of the classical linear cortical
magnification factor. Therefore, we will use $M$ without any index for linear
magnification in this direction. It is convenient to plot the inverse linear
cortical magnification factor, because this is a linear function in $\theta$ on
HM in the monopole map. As the meridians in the monopole map change from
horizontal to vertical, the inverse linear cortical magnification factor grows
slower than linear (see Fig.~3).  This is readily understood, considering the
layout of the visual hemidisc mapped by Eq.\ \ref{eq:logzp1} into its neuronal
representation (Fig.\ 2 {\sf B}).  While the representations of all meridians
extend approximately equally long into the cortical $u$-direction (in
anatomical terms the posterior-anterior axis), the more vertical they are, the
larger is their evasion into the cortical $v$-direction (into the dorsal and
ventral direction for the upper and lower visual quadrant, respectively).  The
increase in cortical space on the vertical meridian (VM) as compared to HM is
about 20\%.  While in fact one study indicates the opposite asymmetry with HM
being cortically over respresented with respect to the vertical meridian
\cite{Janik2003}.

\begin{figure*}[!tpb]
\centerline{\includegraphics[width=\textwidth]{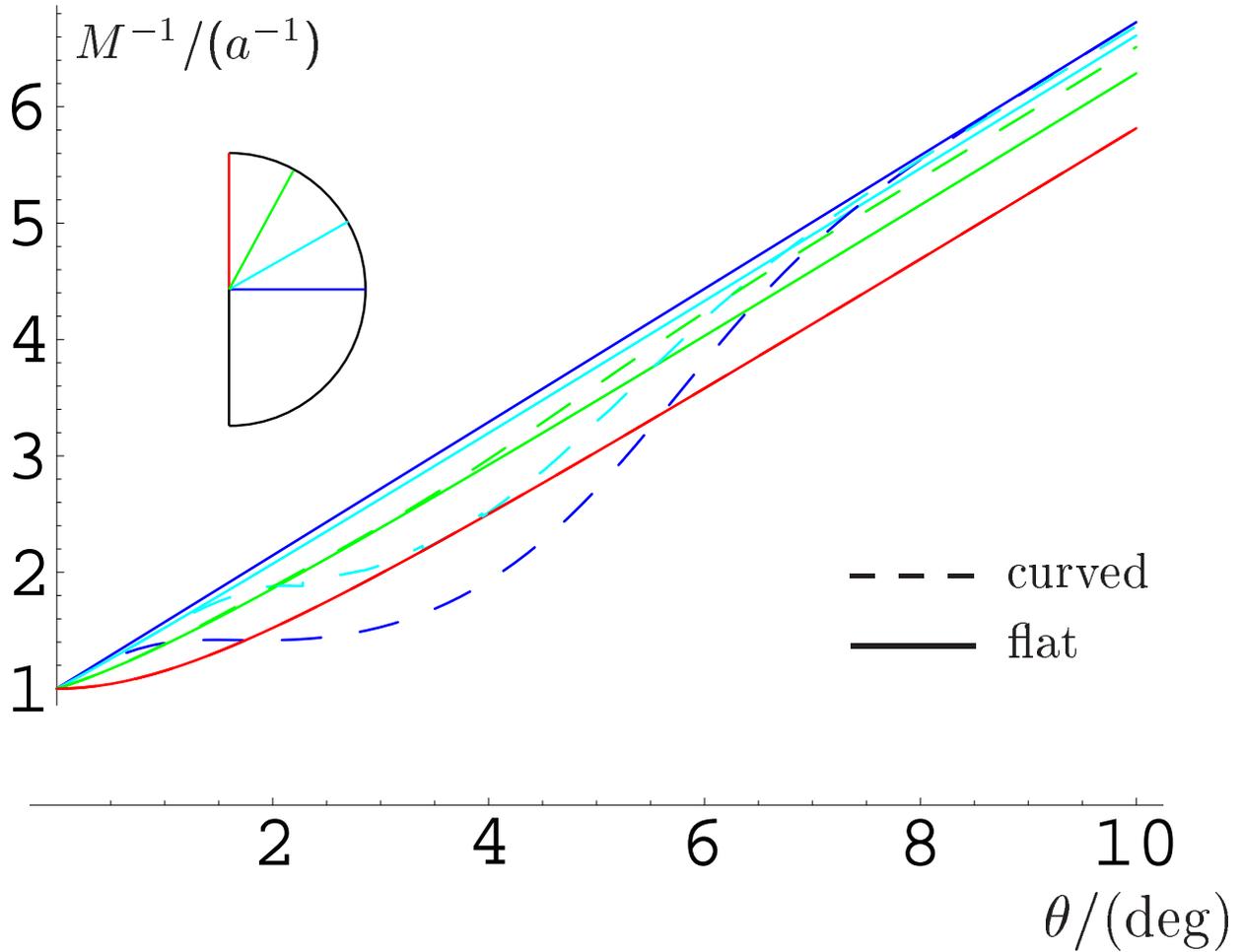}} 
\caption{ 
Inverse linear magnification factor as a function of eccentricity $\theta$
along different meridians (horizontal meridian (HM) 0$^\circ$: blue;
30$^\circ$: cyan; 60$^\circ$: green; vertical meridian (VM) 90$^\circ$: red,
see inset).  Solid lines are on the flat model, dashed on the curved model of
V1. $M$ has an inverse linear dependence on retinal eccentricity $\theta$ only
along the neural representation of HM on the flat model of V1 (solid blue
line).  While on the flat model of V1, $M$ increases with azimuthal distance at
constant eccentricity (solid lines). This meridional asymmetry is reversed on
the curved model of V1 (dashed lines). In particular meridians close to HM gain
cortical surface on the 3D model of V1.  \label{fig:inversM}} 
\end{figure*}


\subsection{3D model of V1 with curved retinotopy}

We construct the  3D model of V1 based on our own and published data obtained
at autopsy of neurologically normal human brain
\cite{Stensaas1974,Andrews1997}. The dimensions of V1 were determined from
cross-sections in both hemispheres. The surface of V1 can easily be identified
postmortem by the stria of Gennari.  This band of myelinated axons can be
traced (see Fig.~1 {\sf G}-{\sf H}).  Such traced curves provide lift functions
$\lift(u,v)$ for forming the originally flat monopole map (Eq.\
\ref{eq:logzp1}) into its 3D from.   We will only  summarize the data obtained
by visual inspection of the stria of Gennari used to define plausible functions
that define the 3D form.  In particular, we outline the restrictions we impose
regarding plausible symmetry principles.

First, we transform the complex function in Eq.\ \ref{eq:logzp1} into the real
$\mr^2$ domain
\begin{equation}
\label{eq:uv}
\begin{array}{ccl}
u (\theta, \phi) & = & \frac{1}{2 b}\log \left(\frac{a^2+2 b \theta \cos(\phi) a+b^2 \theta^2}{a^2}\right), \\  
v (\theta, \phi) & = & \frac{1}{b}\arctan \left(\frac{b \theta \sin (\phi)}{a+b \theta \cos (\phi)}\right). \\  
\end{array}
\end{equation}
Eq.\ \ref{eq:uv} describes together with the lift function $\lift(u,v)$ a
$(\theta, \phi)$-parameterized surface in $\mr^3$, that is, in the Cartesian
coordinates $(u,v,l)$. Note that the lift functions are used in a first step to
describe the surface but not the retinotopy on it. {We define lift functions in
such a manner that the initially  flat surface becomes intrinsically curved.
Note that we may introduce a bias since we actually treat the intrinsically
flat surface also as extrinsically flat at this step of the construction. But
the bias does not necessarily affect retinotopy because we rearrange the
location of the meridians.} Applying the lift function directly to the
retinotopic grid (Fig.~2~{\sf B}) would drive some adjacent meridians farther
away from each other than others, depending on $\frac{\partial
\lift(v,u)}{\partial v}$.  Therefore, to obtain a curved retinotopy we need to
rearrange the location of the meridians by a inverse sampling technique, as
will be described at the end of this section.  

V1 is located entirely or nearly entirely on the medial surface of the
occipital lobe. We align parallel to this surface the $(u,v)$-plane and define
the lift functions $\lift(u,v)$ such that  about two-thirds will lie within the
CS walls.  Futhermore, $\lift(u,v)$ depends on the steepness of the walls of CS
and on its course in the $(u,v)$-plane (see Fig.~1 {\sf A}).  Due to the large
variations among individuals, we have to make some simplifications. We assume
that the course of CS can be deformed to follow a straight course.

To finally construct the lift functions $\lift(u,v)$, it is useful to define
another landmark: the fundus of CS (FCS). Roughly speaking, it is the curve of
maximum depth that spans the length of the CS. A more precise definition can be
given based on curvature or based on distance functions \cite{Kao2006}. Based
on FCS, the assumption that the course of CS is straight can be formulated in a
different way. In this case, the FCS assumes a curved line without torsion and
its projection onto the $(u,v)$-plane is a straight line (see Fig.\ 1 {\sf C}).
For simplicity,  we assume that CS has a bilateral symmetry with the symmetry
plane going through this line. Furthermore, the mathematical description of V1
is w.l.o.g.\ simplified, if the projection of FCS onto the $(u,v)$-plane lies
on the $u$-axis.  Then the family of parametric profiles
$\lift_{u}(v)\equiv\lift(u\!\!=\!\!\mbox{const.},v)$ in the coronal planes
$(v,\lift)_{u}$ completely describe the 3D-configuration of V1.

It significantly simplifies the later performed resampling to define the lift
functions along curved coordinate lines in the $(u,v)$-plane with constant
$\theta$, instead of along the straight coordinate lines with constant $v$.
Since the family of parametric profiles $\tilde\lift_{\theta}(\phi)$ live in
curved planes we transform them into a Cartesian  $({\xi_1},{\xi_2})$ plane
(see Fig.\ 1 {\sf F}).  This results in a family of parametric profiles
$\tilde{\tilde{\lift}}_{{\xi_1}}({\xi_2})$, which we finally choose to be
Gaussian-shaped:
\begin{equation}
\label{eq:liftxy}
\tilde{\tilde{\lift}}_{{\xi_1}}({\xi_2})=d\; e^{-\frac{{\xi_2}^2}{{\sigma_{\xi_2}}^2}}. 
\end{equation}
The parameter $\sigma_{\xi_2}$ defines the width of CS. Its value is
$\frac{\pi}{10}$  (see Fig.\ 2 {\sf C}).  The parameter $d$ gives the depth of
FCS as a function of the new eccentricity coordinate ${\xi_1}$ in the
$({\xi_2},l)$-plane.  Again we choose a Gaussian-shaped profile 
\begin{equation}
\label{eq:depth}
d({\xi_1})=\frac{d_{max}}{s} \left(1-e^{-\frac{{\xi_2}^2}{\sigma_{\xi_1}^2}}\right).
\end{equation}
The parameter $\sigma_{\xi_1}$ defines the steepness of CS. Its value is
$\frac{\pi}{50}$ (see Fig.\ 2 {\sf C}). The parameter $d_{max}=170.94$ gives
the maximal depth of CS in units of $a$, equivalent to 20mm.  The factor
$s=\frac{1}{\pi}\int^\frac{\pi}{2}_{-\frac{\pi}{2}}\sqrt{1 + \left(\frac{d u}{d
\theta}\right)^2} d \theta$  is needed to resize the profiles, since we define
the lift function on the $({\xi_1},{\xi_2})$ plane, but want to have a
Gaussian-shaped FCS in the $(u,l)$ plane.  Each arc with constant eccentricity
spans an angle $\pi$ in the visual field. On the $({\xi_1},{\xi_2})$ plane
these arcs have all the same length, while on the $(u,v)$-plane their arc
length is $\pi\,s$. Therefore, we have to resize $\lift_{{\xi_1}}({\xi_2})$ by
this quotient.  

The Gaussian curvature $K$ is determined by this constrution because it depends
only on the metric (the retinotopy depends on the metric in a retinal
coordinate systems).  In Fig.~4 we show $K$ in a color code on this surface. It
is nearly everywhere close to zero, except for two main locations. One is at
the saddle-shaped entrance of the FC. Here, K is negative (blue). Father along
the fundus, when CS reaches its maximal depth K changes to positive values
(red).

\begin{figure*}[!tpb]
\centerline{\includegraphics[width=\textwidth]{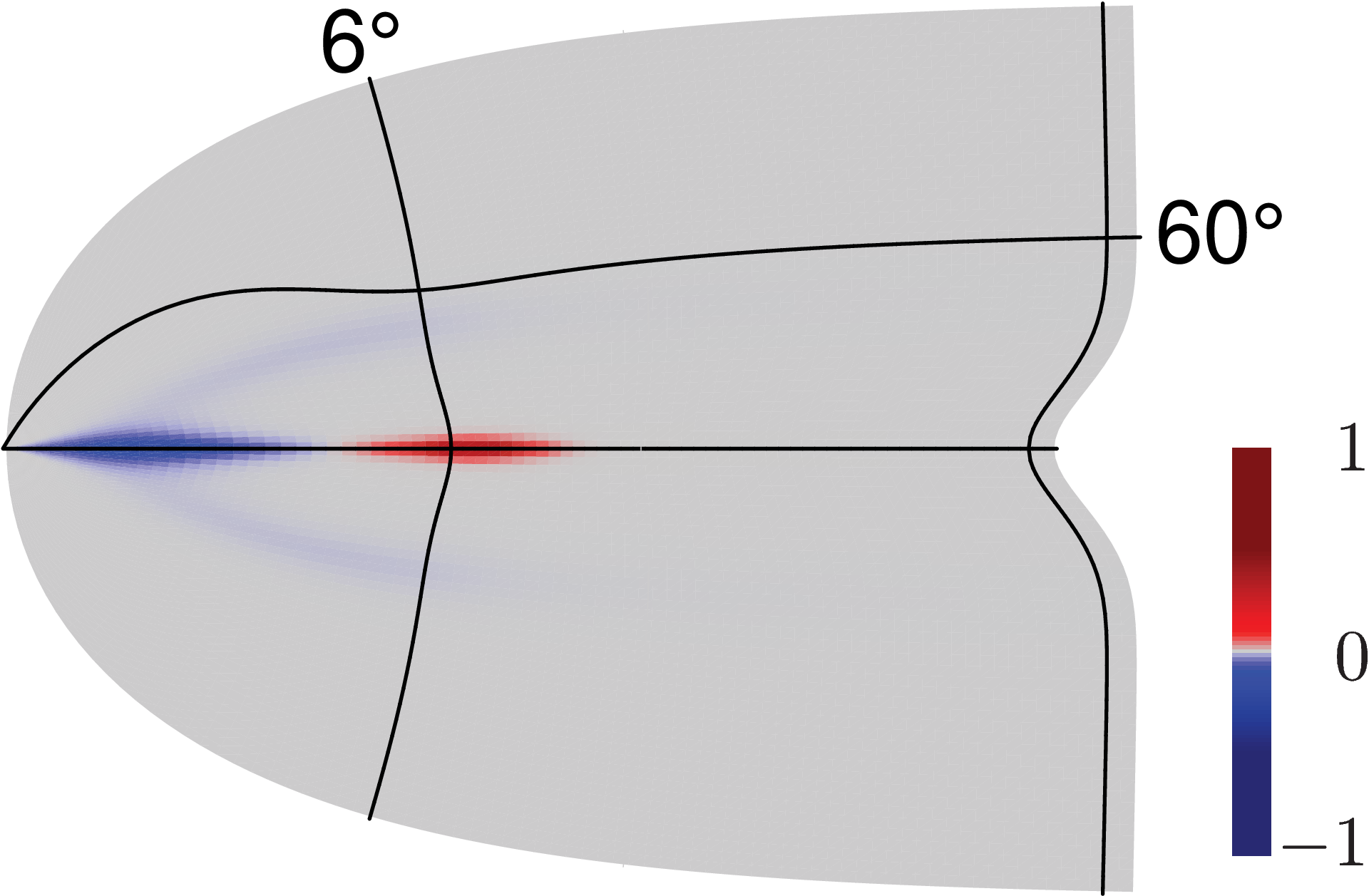}}
\caption{The Gaussian curvature on the curved model of V1 (cf. Fig.~2C)  
}\label{fig:gCurvature}
\end{figure*}

To summarize, four hypotheses have been make about the area V1: (i) two-thirds
will lie within the CS walls, (ii) the walls are symmetric with respect to the
fundus, (iii) the shape of the walls can be approximated with smooth
Gaussian-shaped profiles with constant standard deviation, and (iv) the shape
of the fundus can also be approximated with smooth half Gaussian-shaped profile
(we have alternatively chosen cosine-shaped profiles for both walls and fundus,
with almost identical results) 

\subsection{Retinotopy on a 3D model of V1}

To compare retinotopy on the flat and curved model of V1 we now define the
retinotopic  grid on the latter.  Although this may seem obvious, we want to
note, that we actually define the retinotopy on the curved 3D surface by this.  

Firstly, assume that HM maps to the fundus of the curved model surface.  By
definition HM splits the retinotopic map into two quarter fields, a dorsal
($v\!>\!0$), lower field ($\phi\!<\!0$) and a ventral ($v\!<\!0$), upper field
($\phi\!>\!0$).  If we simply lift the neural representations of the meridians
from the flat surface model of V1 onto the dorsal and ventral part of the
curved model by Eq.~\ref{eq:liftxy}, the meridians drive away from each other
in proportion to $\frac{\partial \lift(v,u)}{\partial v}$.  In fact, meridians
should by equidistantly spaced in the curved model of V1, as they are on the
flat surface (see red coordinate lines in Fig.~2 {\sf C}). This can be achieved
by rearranging $\phi$ on the flat model of V1 before lifting with the inverse
sampling equation
\begin{equation}
\label{eq:iSampling}
\phi(\tilde\phi) = \ell_\theta^{-1}\left(  \frac{2 \tilde\phi -
  \pi}{\pi} \ell_\theta\left(\frac{\pi}{2}\right)\right).
\end{equation}

The function $\ell_\theta(\phi)$ gives the arc length of the profile
$\lift_{\theta}(\phi)$ between the origin ($\phi\!=\!0$) and the value of the
argument $\phi$. $\ell_\theta^{-1}$ is the inverse function with the new value
of $\phi$ as the return value. The rearranged meridians are then lifted down
from the flat model with lift function (Eq.\ \ref{eq:liftxy}).  At last, we
need to place the iso-eccentricity coordinates, i.\,e., the green (inner
10$^\circ$) and blue (10$^\circ$--90$^\circ$) coordinate lines in Fig.~2~{\sf
C}.  They are lifted from the flat model with the lift function (Eq.\
\ref{eq:liftxy}) without rearrangement.

\subsection{Areal magnification and surface area gain}
\label{sec:amf}

Figure 5 {\sf A} shows the areal cortical magnification factor projected in the
visual hemifield for the curved V1. This factor predicts the retinal ganglion
cell density, assuming this quantity is $M$-scaled \cite{Wassle1989}. There is
a larger value of $\det\cmt$ on a stripe centered around HM.  The inverse
profile of M along this stripe is visualized in Fig.~3.  The graph of $M$ in
Fig.~5 {\sf B} visualizes this stripe as a shoulder in the exponentially
decreasing curve $M(\theta)$ at HM.  The shape and location of this stripe in
the visual field (see Fig.~5 {\sf A}) is similar to a retinal specialization
known as visual streak. The visual streak is a stripe of elevated neuronal
density along HM in the retina.  It is found in some animals
\cite{Steinberg1973,Picanco-Diniz1991,Peichl1992,Vidyasagar1992,Guo2000,Arrese2003,Calderone2003}
but  it is not pronounced in humans \cite{Stone1981,Wassle1989,Curcio1990}. The
stripe centered around HM can be thought of as a virtual visual streak as
described in the following. 

The neuronal density in the retina is assumed to be M-scaled (areal
magnification).  Let us define a virtual counterpart of the visual streak as
the elevation of inversely retinotopic projected densities of cortical cells.
Along the horizontal meridian, the retinal representation of cortical cells is
largely elevated with respect to other more vertical meridians outside this
visual streak.  A virtual visual streak is related to surface area
magnification but also to surface gain by folding.  Just as Eq.\ \ref{eq:norm}
defined the stretch of a unity vector, we can calculate the surface area gain
of a unity area in the retina via the expansion tensor $\vecBold{C}$.
Considering rigid body transformations, $\vecBold{C}$ is the unity tensor
$\vecBold{1}$ and surface area gain is a measure that must be zero.  This led
us to define the strain tensor $\vecBold{\varepsilon}=\frac{1}{2}(\vecBold{C} -
\vecBold{1})$.  In continuum mechanics $\vecBold{\varepsilon}$ is called the
Green-Lagrange strain tensor.  The tensor $\vecBold{\varepsilon}$ is a measure
of surface gain from retina to cortex. The measure of surface area gain by
cortical folding is obtained when the retinal reference frame is replaced by
its neural representation on the flat model surface of V1.  The strain tensor
of the curved model of V1, with the cortical flat reference frame, is  called
$\vecBold{E}$. It measures the surface gain by folding.

The value of $\det \vecBold{E}$ corresponds to the elevation of retinal density
that is needed to satisfy $M$-scaling in a folded cortex. To show that this is
in good agreement with some experimental data \cite{Lashley1941,Sakaguchi2003}
(see also discussion), we look at $\det \vecBold{E}$ along different meridians
(Fig.~5 {\sf C}). For all meridians, folding provides equally increased
cortical space for the representation of the periphery of the visual field.
This increase is mainly limited to the periphery because the calcarine sulcus
just starts at the neural representation of the fovea and additional surface
area is mainly gained when the calcarine sulcus has reached its maximum depth.
Close to the fovea, meridians close to HM also gain  cortical surface area for
their neural representation (see peak in Fig.~5 {\sf C} at HM).

The location of the virtual visual streak, i.\,e., increased values of $\det
\vecBold{E}$, depends on the particular form of V1. Our 3D surface model of V1
averages over the observed substantial variability
\cite{Stensaas1974,Gilissen1996,Andrews1997}. A statement about the location of
the virtual visual streak independent of this variability can be done when we
describe the location of the virtual visual streak relative to the curvature of
V1.  For this purpose one needs to consider the Gaussian curvature, because it
is determined only by intrinsic properties of the surface. The Gaussian
curvature of a point on a surface is the product of the maximal and minimal
curvature of all curves passing through this point.  As a consequence, a
saddle-shaped surface has negative Gaussian curvature while a spherical surface
is positively curved.  The entrance of CS is saddle-shaped and has therefore
negative Gaussian curvature. The location where the fundus of the calcarine
sulcus reaches its maximal depth has positive Gaussian curvature.  The peak of
the virtual visual streak is located at the transition between negative and
positive Gaussian-curved areas (Fig. 5 {\sf D}).  While this provides only a
rough estimate of the location of a virtual visual streak it is not only
independent from the variability but can also be used as a prediction for
selective magnification in other areas in the cerebral cortex.

\begin{figure*}[!tpb]
\centerline{\includegraphics[width=\textwidth]{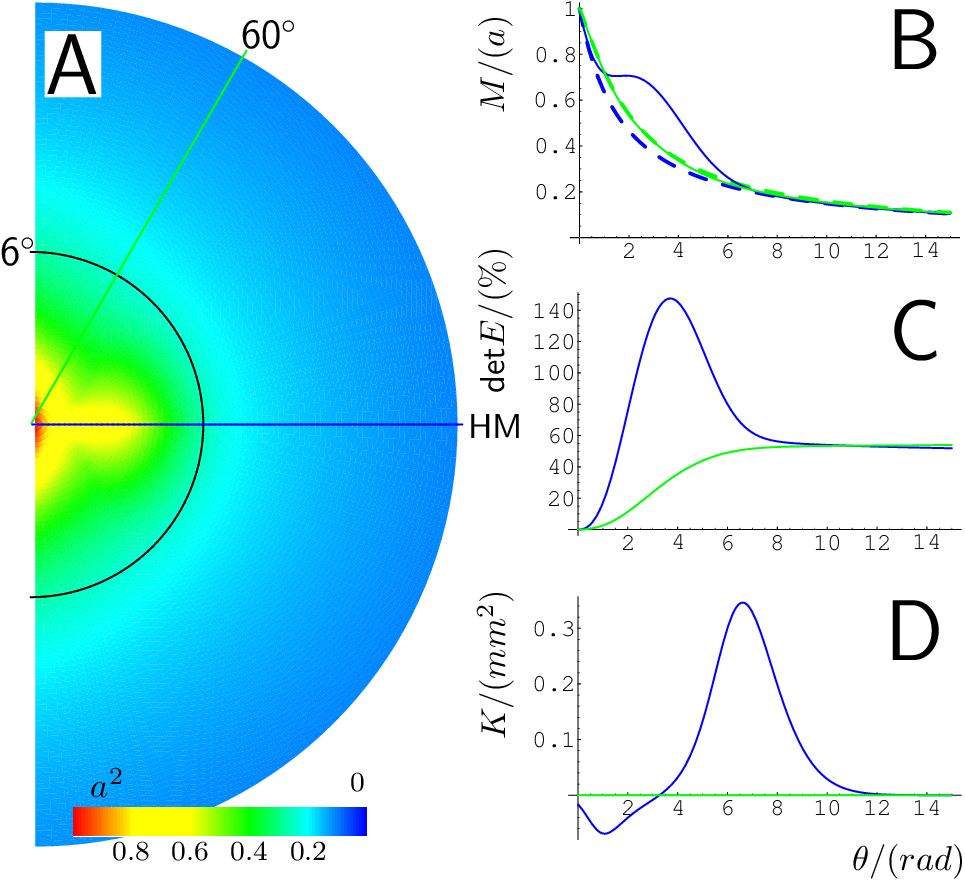}} 
  \caption { 
({\sf A})  Areal cortical magnification factor $\det\cmt$ projected onto the
visual hemifield.  The color bar indicates the normalized value in units of
$a^2$.  The yellow blob at  an eccentricity $\theta=3^\circ$ along the horizon
indicates the virtual visual streak.  ({\sf B}) Linear cortical magnification
$M$ vs eccentricity $\theta$ in units of $a$ along the horizontal meridian at
$\phi=0^\circ$ (HM) and the meridian at $\phi=60^\circ$ (dashed and solid lines
on flat and curved model of V1, respectively).  ({\sf C}) Normalized surface
area gain  $\det \vecBold{E}$ and ({\sf D}) Gaussian curvature  along HM and
$\phi=60^\circ$.  \label{fig:arealMagps}} 
\end{figure*}

\section{Discussion}
\label{sec:discussion}

Cortical magnification is the local description of retinotopy, or in fact, of
receptotopic maps in general.  Tonotopic and somatotopic maps are other
prominent examples \cite{Gaschler-Markefski1997,Rademacher2001,Lotze2000} for
which cortical magnification can be defined in the same manner as for the
visual modality.  Cortical magnification is the most fundamental quantity of
cortical feature maps that describes how much cortex is devoted to process
sensory information.  A thorough mathematical foundation of magnification is
needed, if we want to describe anisotropies and other symmetry breaking
constraints in such maps that may serve distinct purposes in sensory
information processing.   We will start our discussion with  experimental
evidence supporting our predictions, continue with methods able to verify them,
and provide the context to current alternative tensor methods. Then, the
limitation in symmetry properties of  flat models of retinotopy and our
assumptions concerning the 3D surface model of V1 are discussed.  We finish
with conclusions that can be drawn from our results on sulcal development.

\subsection{Experimental studies supporting meridional asymmetry}

The approximate form of human retinotopy has long been established from various
methodologies. Some methods, which were mainly used before non-invasive imaging
became available, obtain $M$ indirectly by assuming a $M$-scaled quantity and
measured this. If $M$ is known at various loci and also additional constraints
are given, such as certain symmetries, one can determine the layout of the
global retinotopic map.  Psychophysical studies that obtain an $M$-scaled
quantity provide an elegant measure of cortical magnification mainly because
they do not require an explicit reconstruction of the cortical surface.  They
greatly simplify the analysis bypassing any potential problems concerned with
surface reconstruction. Despite this advantage, these methods suffer from
limited resolution, poor quantification and are unable to directly measure $M$.
They rely on the assumption of $M$-scaling, a principle that could be violated.

Anyway, evidence for the predicted meridional asymmetry is found in a
study investigating perceptual filling-in \cite{Sakaguchi2003} and in a report
about the shape of travelling migraine scotoma \cite{Lashley1941}.  Both
studies indicate that cortical magnification is not meridional asymmetric but
selectively increased on or close to horizontal meridian (HM). 

The perceptual filling-in refers to the tendency of stabilized retinal stimuli
to fade and become replaced by their background. The time required for this
illusion was investigated \cite{Sakaguchi2003}.  It varies with the azimuthal
angle $\phi$ of the target in the visual field. The filling-in was facilitated
as the target position changed from the horizontal to the vertical meridian.
The time was maximal in the horizontal direction and minimal in the vertical.
The absolute value differed in the several experiments, which were performed to
exclude artifacts. But in general for a target presented at 8$^\circ$
eccentricity the response time for the horizontal condition was more than 30\%
longer than the one for the vertical.  The author was aware of the fact that
his findings can be explained by anisotropic cortical magnification. If
magnification on HM is selectively increased, the neural representation of
stimuli located on HM is increased too, and the time required for perceptual
filling-in is prolonged.

\begin{figure*}[!tpb]
\centerline{\includegraphics[width=\textwidth]{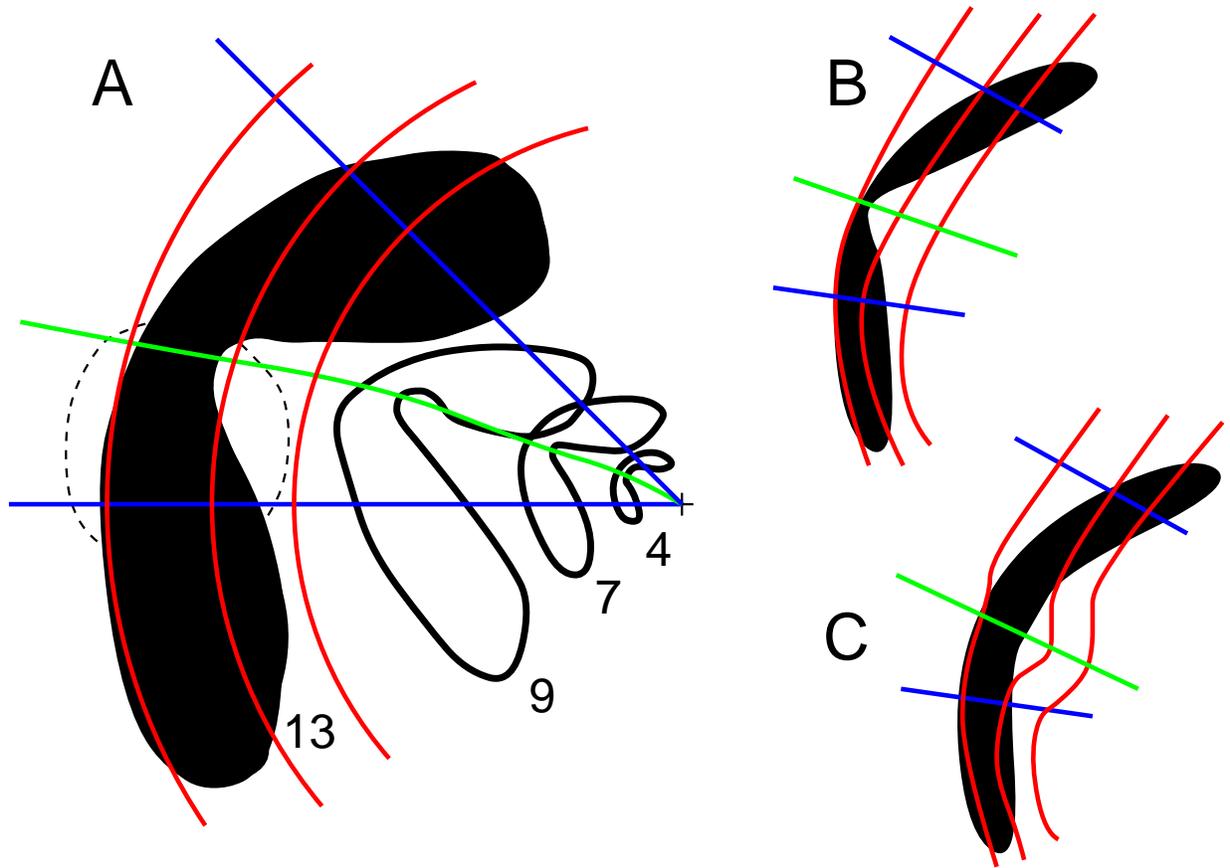}} 
\caption{
({\sf A}) Four subsequent snapshots of a travelling migraine scotoma drawn by
Lashley \cite{Lashley1941}. The small cross at the left indicates the center of
gaze, and the numbers state the time in minutes elapsed before the location of
the scotoma was outlined.  The point of time when the neurological symptoms
were first noticed was  chosen as the start. The scotoma travels across the
left visual hemifield. ({\sf B}) and ({\sf C}) Two different neural patterns in
V1 that can explain the scotoma in ({\sf A}).  Either the noticeable notch in
the scotoma pattern along the green path in ({\sf A}) is reflected in the
neural pattern by a narrow wave width along the neural representation of the
green path ({\sf B}) or the retinotopic map is distorted and the wave width is
constant ({\sf C}). In this case, the green path marks the fundus of the
calcarine sulcus.\label{fig:lashleyLVF}}
\end{figure*}

Similar support for selectively increased cortical magnification on HM comes
from the shape of travelling scotomta observed during a migraine attack with
aura (Fig.\ 6 {\sf A}). One of the first scientific reports on migraine with
aura { was given by Lashley \cite{Lashley1941}. He } described in astonishing
detail the spatio-temporal patterns of travelling scotoma in his visual field
during migraine.  The cortical magnification was not considered directly. But
the shape of the scotoma in his left visual hemifield clearly hints to
meridional asymmetry in $M$.  Figure 6 {\sf A} reproduces the spatio-temporal
patterns drawn by Lashley.  For the purposes of illustration it was modified: a
polar grid was added,  the first marked position of the scotoma is not shown,
and  the latest observed loci of the scotoma is shown as a curve being filled
black.  This particular shape and location was observed about 13 minutes after
the onset of the attack. Two possibly underlying neural patterns are shown in
Figure 6 {\sf B} and {\sf C}. Assuming a constant speed of 2.5~mm/min of the
pathophysiological process within V1, the scotoma at minute 13 (filled) is
about 32.5~mm away from the neural representation of the fovea. Mapped with
Eq.\ \ref{eq:logzp1} and the parameter values $a=0.117$ and $b=0.067$ this
would correspond to 13.6$^\circ$ degrees eccentricity. This is in sufficient
agreement with the position of the retinal blind spot marked by the dashed
circular curve.

Independent of the parameter values for $a$ and $b$ of the monopole map in
Eq.~(\ref{eq:logzp1}), the scotoma shown in Figure 6 {\sf A} cannot be mapped
such that its gross shape changes. In other words, the narrow notch  along the
green path will not disappear.  Therefore, either the width of the underlying
cortical wave (a spreading depression \cite{Dahlem2003,DAH08d}) varies along
its path in the cortex (Fig.~6 {\sf B}) or, much more likely, retinotopy is
distorted from that of the monopole map (Fig.~6 {\sf C}). The latter is in
agreement with our predicted selective increase of cortical magnification along
the visual location of the steep fundus of the calcarine sulcus (green path).
This provides a test for our 3D model retinotopy. When the selectively
increased magnification follows a path not exactly on HM but is shifted to the
upper (or lower) field, the neural representation of this path should have the
largest length from the neural representation of the fovea to most anterior
part of the retinotopic map. If this path mainly overlaps with the steep
fundus, the cortical representation of the horizontal meridian is  shifted to
the ventral (or dorsal) bank of calcarine sulcus.  A qualitative analysis shows
that one can measure linear cortical magnification precisely by comparable but
simpler drawings of visual migraine aura \cite{Gruesser1995}.  Unfortunately
this data was pooled for all meridional directions.

\subsection{Alternative tensor methods}

By definition, methods that obtain $M$ indirectly suffer from the possibility
that the assumption of $M$-scaling locally fails.  Only direct methods that
precisely delineate the retinotopic organization of human visual can measure
cortical magnification.  These technique are mainly used today to model
structural and functional variation of the human brain. They can detect group
differences, for example, disease effects on cortical anatomy
\cite{Baseler2002,VanEssen2006}, but also $M$ was investigated
\cite{Sereno1995,Engel1997,Duncan2003}.  Problems occur in these studies by
smoothing and flattening experimental data, because these techniques are based
on mathematical models of retinotopy that introduce certain symmetries (see
next Section).  A recent approach to estimate $M$ \cite{Qiu2006} avoids many of
the these problems because it is completely data-driven.  It preserves geodesic
distances by directly computing $M$ on the folded cortical surface.  They found
that $M$ is similar in the dorsal and ventral compartments of V1 within each
hemisphere.  A selective analysis of $M$ on HM is not included.

These direct methods to obtain $M$ also use a tensor approach to morphometry.
They are mainly related to the Laplace-Beltrami operator. One may ask: How do
these approaches compare to our tensor description?  

In these studies the specific cortical surface parameterization is not
essential, though much effort has gone in the development of computationally
advantageous parameterizations.  If $\vecBold{g}_{V1}$ denotes the cortical
metric in an arbitrary parameterization, its components can be used to define
the Laplace-Beltrami operator as a generalization of the Laplacian differential
operator. In contrast, to derive the cortical magnification tensor \vecBold{M} the
parameterization is essential.  The cortical surface parameterization by
sensory coordinates is, in fact, the very idea behind cortical magnification.
That is why we have introduced the term cortical expansion tensor $\vecBold{C}$
for this particular form of $\vecBold{g}_{V1}$.  For other tensor approaches it
is natural and often convenient for the purposes of visualization to
parameterize the cortex after inflating it to a sphere using the induced
spherical coordinates.  One has to define sulcal curves as anchors, e.g., the
Talairach coordinate system \cite{Talairach}, to achieve that corresponding
anatomical landmarks, such as the very cortical sulci, occur at the same
spherical coordinate across subjects.  The Lagrangian formulation of a
receptotopic map is a tensor approach that provides a natural and independent
parameterization of the cortical surface.  According to this description data
is pooled across subjects in the sensory reference frame, which sort of
replaces anatomical anchors by functional labels.  A similar idea is used by
individual-subject region of interest analysis to identify functional labels
\cite{Kanwisher1997,Epstein1998,Downing2005}.

\subsection{Influence of ocular dominance}

The wedge-dipole model of retinotopy \cite{Balasubramanian2002} introduces two
extensions two the monopole map. It is firstly introduced to model multiple
fields (V1-V2-V3 complex) and also to improve the representation of peripheral
data.  The term ``{\em dipole}'' in its name refers to the latter, which is,
however, for our study rather irrelevant.  The corresponding mathematical
changes that also introduce additional parameters are therefore not further
considered here. We will only discuss the other extension referred to by the
term ``{\em wedge}''. In mathematical terms wedge refers to a methods that
compresses the visual hemifield along iso-eccentricity lines
($\theta\!\!=\!\!\mbox{constant}$) to a wedge with apex angle below $\pi$.
Usually, it is a half disc (apex angle equals $\pi/2$).  This extension can
actually serve two purposes.  Firstly, but as mentioned not relevant in the
context of this study, the compression allows a unified model for the
topography of the full visual field in areas V1, V2, and V3.  Secondly, the
compression induces a topographic shear.  Such a shear was stated to relate to
the combination of two full representations of the visual field into V1 by
ocular dominance columns \cite{Tootell1982,Sakitt1982}.  The orientation of the
stripes, in which ocular dominance organize and which define the direction of
the topographic shear, may well relate to the 3D form of V1.

In primates the patterns of segregated ocular dominance stripes run mostly
parallel to the neural representations of $\theta$ coordinate lines (for the
pattern in human see \cite{Horton1984}). At least for non-human primates in a
region close to the fovea stripes tend to run horizontally \cite{LeVay1985}.
In the direction perpendicular to the stripes, twice as much cortical space is
needed to inject the two ocular representations of the visual field. Therefore,
$M$ depends on direction.  For the sake of simplicity, consider the patterns of
ocular dominance stripes runing parallel to $\theta$ coordinate lines in the
whole visual field.  $M$ is usually measured along the $\theta$ direction
perpendicular to the ocular dominance stripes. The value of $M$ should be only
half as large in opposite direction parallel to the ocular dominance stripes,
because only one visual hemifield is mapped along this direction.  The effect
of such a constant rotational shear, which is not accounted for in our model,
is just a scale factor of $\frac{1}{2}$. For example the flat and curved model
of V1 would be half as wide. Since this induces a quasi-conformal map on the
flat surface model the symmetry arguments given above are still valid except
for the scaling. However, in the foveal region stripes tend to run
horizontally. The opposite scaling effect is obtained when $M$ is measured only
in this region. If data for $M$ is obtained as an average from both regions the
effects can partly cancel.  A detailed analysis of the effect of dominance
stripes on cortical magnification and cortical curvature is on open problem.

\subsection{3D model of V1 and its parameterization}

We have made four partly interlinked hypotheses to construct the averaged 3D
form of V1, namely that (i) two-thirds of the surface area lie within the CS
walls,  (ii) the ventral and dorsal half of V1 are symmetric with respect to
the fundus, in particular, they extend equally long into the anterior
direction,  (iii) the shape of the walls can be approximated with smooth
profiles (Gaussian-shaped profiles with constant standard deviation) and (iv)
the shape of the fundus can also be approximated with smooth a half
Gaussian-shaped profile.  To obtain a curved retinotopic map on this surface
area of V1, three further hypotheses are made: namely, that (v) the horizontal
meridian maps to fundus of the calcarine sulcus, and (vi) that meridians are
equidistantly spaced in the curved V1 (Eq.\ \ref{eq:iSampling}), and (vii) in
all other respects the retinotopy is goverend by the original monopole map.

The  hypotheses (i)-(iv) are, to our mind, justified because some smaller
deformations within the gross shape of the calcarine sulcus are probably not
systematic and we assume that, on average, their influence is small compared to
the curvature effects we investigate in this study. The course of calcarine
sulcus varies widely among individuals and the lunate sulcus, which is roughly
oriented vertically to it, appears to mark the anterior boundary of V1 in
non-human primates, but it is often missing in humans \cite{Connolly1950}.
Autopsy data  suggest that V1 proceeds farther anteriorly in the lingual gyrus
\cite{Stensaas1974,Andrews1997}, while fMRI data show that the dorsal and
ventral compartments of V1 are at least similar in absolute extent
\cite{Qiu2006}.   If V1 proceeds farther anteriorly in the lingual gyrus this
can be balanced by higher curvature of this gyrus. 

Our main results are robust to some small deviations of these
hypotheses, because we only investigate the influence of the gross sulcal
pattern on a standard retinotopic layout. For example, we have replaced the
Gaussian-shaped profiles (iii,iv) with cosine-shaped profiles with similar
results.  Furthermore, if (v) holds only approximately and a meridian in the
lower or upper quadrant of the visual hemifield with a mild negative or
positive shift away from horizontal meridian, respectively, maps onto the
fundus, cf.~\cite{Aine96}, we can still draw our conclusions as to where is the
virtual visual streak.  In this case, the retinotopy and in particular its
derivation $M$ can still be rather similar in the dorsal and ventral
compartments of V1, because hypotheses (v) and (ii) are interlinked and the
ventral and dorsal half of V1 (ii) can shift accordingly to respect the major
symmetry between upper and lower visual quarter fields \cite{Qiu2006}.

In fact, it provides a test to verify the predicted correlation between
anatomical landmarks and functional properties if (v) does not hold exactly and
instead of the horizontal meridian another path in the visual field, possibly
with a variable azimuthal offset with respect to the horizon, maps onto the
fundus. Suppose, for example, that a meridional path with a positive shift away
from the horizon (upper visual quadrant) is found with an increased cortical
magnification, as suggested by data shown in Figure 6. Then the neural
representation of horizontal meridian should be on the ventral bank, because
according to our prediction this path is the location of the fundus in the
visual field (if reversed retinotopically mapped).  Overall, these hypotheses
mainly establish a symmetry between the neural representation of the upper and
lower visual field quadrant, as found in \cite{Qiu2006}.

While the 3D form of V1 affects the intrinsic properties of the surface, the
layout of the retinal grid on this curved surface ultimately determines \vecBold{M}.
It was been chosen in a somewhat teleological manner { by enforcing that
meridians are equidistantly spaced (vi) while at the same time providing more
cortex along the fundus that is created by the fold.}   Although this seems
reasonable, one should notice that the location of the virtual visual streak
depends on this assumption.  Furthermore, because of this assumption, \vecBold{M} is
not sensitive to the precise shape of the profiles of the lift function {(in
particular the Gaussian-shaped profiles with constant standard deviation
(iii))}.  An alternative and completely self-organized method can be provided
by neural networks such as a Kohonen net. To model retinotopy  on the curved
V1, the 2D net of the Kohonen layer must represent in its lateral connections
the metric of the curved V1.  The natural learning set is the cell density in
the retina. We have investigated such networks and found similar results
\cite{tusch04}. The major difference is that the increased areal magnification
on the horizontal meridian is shifted slightly into the periphery, where the 3D
form of V1 has positive Gaussian curvature.  Whereas the maximum areal
magnification in this analytic study is found in between the change from
negative to positive Gaussian curvature (Fig.~5 {\sf D}). 

\subsection{Development of cerebral sulci}

On functional grounds, an increased cortical representation of the horizon with
respect to the average cortical magnification on meridians (Fig.~5 {\sf A})
serves a similar purpose as a visual streak.  It is important to note that our
model predicts where a virtual visual streak is located, while the existence of
a virtual visual streak is a mathematical consequence of the fact that a
surface with non-constant Gaussian curvature cannot be isometrically embedded
in one with constant Gaussian curvature (e.g. intrinsically flat surface).
Hence, the purpose of our tensor approach is to predict the {\em shape} and
{\em location} of a virtual visual streak within V1. These are not intrinsic
properties of the cortical surface alone but also depend on retinotopy.  The
virtual visual streak therefore provides a link between cerebral sulci and
encephalization of function because it links anatomy to retinotopy.  The
virtual visual streak is defined for the visual modality as a selective
increase of the number of cortical neurons mapped into the visual domain. It
can in principal be defined more generally for all senses and also for motor
function. Such a quantity requires, as cortical magnification does, a notion of
distance in the sensory or motoric space. For the sensory modalities like
visual, auditory, and tactile, the distance between stimuli can be defined
based on retina, basilar membrane, and body surface, respectively.  The
receptor densities must also be considered ($\det \vecBold{E}$ vs. $\det
\vecBold{M}$). For the sensory modalities like olfactory or also for kinetic
modalities this concept is not readily clear.  For example, given two odorants,
is it possible to define a notion of distance between them?  There is a
stereotyped map of odorant receptor inputs in the olfactory bulb
\cite{Buck2004} but the concept of magnification requires two such maps.
Nevertheless, we believe our approach can be applied to all modalities.  It may
even guide us in the search for a notion of sensory distance for modalities
where it is not naturally  given.

Animals with a real visual streak have lower encephalization than human, even
with rather lissencephalic cerebral hemispheres
\cite{Steinberg1973,Picanco-Diniz1991,Peichl1992,Vidyasagar1992,Guo2000,Arrese2003,Calderone2003}.
When a retina has developed a visual streak, it is not possible to map visual
input spatially homogeneous onto a functional field of a lissencephalic cortex
without changing the global layout of the map.  In other words, the $M$-scaling
principle is likely to be locally violated. In such a case (e.\,g.\ macropus
eugenii), a correlation between retinal ganglion cell density and $M$ was found
only along VM but not HM \cite{Vidyasagar1992}.  Hence, neural input is
increased at the cortical site of the representation of the visual streak. A
surface undulation can provide more cortical surface area for this increased
input and thereby reestablishing the $M$-scaling principle.  Dasyprocta
leporina, another mammalian, in which $M$ along HM approximately corresponds to
its visual streak, shows gradual encephalization of the visual streak.  Its
retinotopy has nevertheless asymmetries that are not directly related to the
topography of the retinal ganglion cell density \cite{Picanco-Diniz1991}.

Irrespective of phylogenetic relationships, species that inhabit open spaces
may independently have evolved a visual streak or a virtual visual streak that
enables them to better scan the horizon. In other words, if we approach human
retinotopy from the comparative standpoint, we can explain why the sulcal
position is where it is, though not necessarily  how sulci develop.  The
specific location of CS might be a convergent evolutionary solution to enhance
visual function on HM. { The human retina poses a very mild visual streak
\cite{Stone1981,Curcio1990} and even without taking the visual streak into
account, the $M$-scaling is violated in all current theoretical models of human
retinotopic maps, because they have a decreased cortical representation of the
horizon with respect to the average cortical magnification on meridians.}  At
the location where $M$-scaling is violated either less or more cortical surface
area is available to process information from a certain number of retinal
ganglion cells.  Therefore, { our predicted selective increase of cortical
magnification due to cortical folding, i.e.,} the virtual visual streak with
selectively increased $\det \vecBold{E}$, indicates a form of encephalized
visual magnification.  If the virtual visual streak is actually a phylogenetic
shadow of a visual streak in human ancestors, {which is supported by its mild
occurrence \cite{Stone1981,Curcio1990},} thus it is not an independent
evolutionary solution, a switch from selectively increased retinal to cortical
magnification on HM took place.  

To summarize, the theme from structure to function is central to biology, as is
the reverse direction from function to structure. We doubt that the precise
correlation between the primary visual cortex, as an functional area, and the
calcarine sulcus, as an anatomical structure, occurred by chance. 

None of the existing mathematical models is suited to describe retinotopy in
relation to the cortical geometry. Most of them are in fact either 1D functions
assuming invers linear mapping given by two parameters ($a$,$b$ or $A$,$E_2$,
see Eq.~\ref{eq:logzp1}), without taking into account that a 2D generalization
is not trivial, or they are intrinsically flat maps described by an analytical
function. The monopole map and its five parameter extension, called
wedge-dipole map \cite{Balasubramanian2002} are by definition conformal and
quasi-conformal, respectively.  Any analytic function has a high inner
structure, i.\,e., exhibit certain symmetries. Alternative nonconformal maps
suffer from inconsistencies with the definition $M$ \cite{DayanAbbott}.  It is
assumed that single components of the Jacobian matrix $\vecBold{J}_\Phi$ (Eq.\
\ref{eq:jacobian}) scale with $M$. But $\vecBold{J}_\Phi$, unlike \vecBold{M},
contains a rotation carrying the retinal directions onto its neural
representations. This renders the assumption implausible if conformal mapping
is not presumed in the first place. 

Generally, when cortical functions became encephalized and even completely new
sensory and motoric areas simultaneously arose, this constitutes a
developmental force for the prominent patterns in the highly convoluted surface
of the cerebral cortex in human. Linking cortical magnification to anatomy with
methods of continuum mechanics and complex analysis may shed some light on what
could have been the underlying pattern formation principles.

\section*{Acknowledgement} We likes to thank Jochem Rieger for providing many
helpful suggestion and Leo Peichl for his  comments on the visual streak. We
also thank one anonymous reviewer for various helpful comments on the manuscript.
JT was supported by a grant of the Deutsche Forschungsgemeinschaft
(DA-602/1-1); MAD acknowledges support by Deutsche Forschungsgemeinschaft in
the framework of SFB910. An revised version was done while MD has been
supported by the Mathematical BiosciencesInstitute at the Ohio State University
and the National Science Foundation under Grant No. DMS 0931642.



{\ifthenelse{\boolean{publ}}{\footnotesize}{\small}
 


\newcommand{\BMCxmlcomment}[1]{}

\BMCxmlcomment{

<refgrp>

<bibl id="B1">
  <title><p>{Cytoarchitectonic definition of prefrontal areas in the normal
  human cortex: II. Variability in locations of areas 9 and 46 and relationship
  to the Talairach Coordinate System}</p></title>
  <aug>
    <au><snm>Rajkowska</snm><fnm>G</fnm></au>
    <au><snm>Goldman Rakic</snm><fnm>P S</fnm></au>
  </aug>
  <source>Cereb Cortex</source>
  <pubdate>1995</pubdate>
  <volume>5</volume>
  <issue>4</issue>
  <fpage>323</fpage>
  <lpage>-337</lpage>
</bibl>

<bibl id="B2">
  <title><p>{{T}hree-dimensional statistical analysis of sulcal variability in
  the human brain}</p></title>
  <aug>
    <au><snm>Thompson</snm><fnm>P. M.</fnm></au>
    <au><snm>Schwartz</snm><fnm>C.</fnm></au>
    <au><snm>Lin</snm><fnm>R. T.</fnm></au>
    <au><snm>Khan</snm><fnm>A. A.</fnm></au>
    <au><snm>Toga</snm><fnm>A. W.</fnm></au>
  </aug>
  <source>J. Neurosci.</source>
  <pubdate>1996</pubdate>
  <volume>16</volume>
  <issue>13</issue>
  <fpage>4261</fpage>
  <lpage>-4274</lpage>
</bibl>

<bibl id="B3">
  <title><p>Structural divisions and functional fields in the human cerebral
  cortex.</p></title>
  <aug>
    <au><snm>Roland</snm><fnm>P.E.</fnm></au>
    <au><snm>Zilles</snm><fnm>K.</fnm></au>
  </aug>
  <source>Brain Res Brain Res Rev.</source>
  <pubdate>1998</pubdate>
  <volume>26</volume>
  <fpage>87</fpage>
  <lpage>105</lpage>
</bibl>

<bibl id="B4">
  <title><p>{Broca's region revisited: cytoarchitecture and intersubject
  variability}</p></title>
  <aug>
    <au><snm>Amunts</snm><fnm>K</fnm></au>
    <au><snm>Schleicher</snm><fnm>A</fnm></au>
    <au><snm>Burgel</snm><fnm>U</fnm></au>
    <au><snm>Mohlberg</snm><fnm>H</fnm></au>
    <au><snm>Uylings</snm><fnm>H B</fnm></au>
    <au><snm>Zilles</snm><fnm>K</fnm></au>
  </aug>
  <source>J Comp Neurol</source>
  <pubdate>1999</pubdate>
  <volume>412</volume>
  <issue>2</issue>
  <fpage>319</fpage>
  <lpage>-341</lpage>
</bibl>

<bibl id="B5">
  <title><p>{fMRI evaluation of somatotopic representation in human primary
  motor cortex}</p></title>
  <aug>
    <au><snm>Lotze</snm><fnm>M</fnm></au>
    <au><snm>Erb</snm><fnm>M</fnm></au>
    <au><snm>Flor</snm><fnm>H</fnm></au>
    <au><snm>Huelsmann</snm><fnm>E</fnm></au>
    <au><snm>Godde</snm><fnm>B</fnm></au>
    <au><snm>Grodd</snm><fnm>W</fnm></au>
  </aug>
  <source>Neuroimage</source>
  <pubdate>2000</pubdate>
  <volume>11</volume>
  <issue>5 Pt 1</issue>
  <fpage>473</fpage>
  <lpage>-481</lpage>
</bibl>

<bibl id="B6">
  <title><p>{Statistical methods in functional magnetic resonance imaging with
  respect to nonstationary time-series: auditory cortex activity}</p></title>
  <aug>
    <au><snm>Gaschler Markefski</snm><fnm>B</fnm></au>
    <au><snm>Baumgart</snm><fnm>F</fnm></au>
    <au><snm>Tempelmann</snm><fnm>C</fnm></au>
    <au><snm>Schindler</snm><fnm>F</fnm></au>
    <au><snm>Stiller</snm><fnm>D</fnm></au>
    <au><snm>Heinze</snm><fnm>H J</fnm></au>
    <au><snm>Scheich</snm><fnm>H</fnm></au>
  </aug>
  <source>Magn Reson Med</source>
  <pubdate>1997</pubdate>
  <volume>38</volume>
  <issue>5</issue>
  <fpage>811</fpage>
  <lpage>-820</lpage>
</bibl>

<bibl id="B7">
  <title><p>{Probabilistic mapping and volume measurement of human primary
  auditory cortex}</p></title>
  <aug>
    <au><snm>Rademacher</snm><fnm>J</fnm></au>
    <au><snm>Morosan</snm><fnm>P</fnm></au>
    <au><snm>Schormann</snm><fnm>T</fnm></au>
    <au><snm>Schleicher</snm><fnm>A</fnm></au>
    <au><snm>Werner</snm><fnm>C</fnm></au>
    <au><snm>Freund</snm><fnm>H J</fnm></au>
    <au><snm>Zilles</snm><fnm>K</fnm></au>
  </aug>
  <source>Neuroimage</source>
  <pubdate>2001</pubdate>
  <volume>13</volume>
  <issue>4</issue>
  <fpage>669</fpage>
  <lpage>-683</lpage>
</bibl>

<bibl id="B8">
  <title><p>{The topography and variability of the primary visual cortex in
  man}</p></title>
  <aug>
    <au><snm>Stensaas</snm><fnm>S S</fnm></au>
    <au><snm>Eddington</snm><fnm>D K</fnm></au>
    <au><snm>Dobelle</snm><fnm>W H</fnm></au>
  </aug>
  <source>J Neurosurg</source>
  <pubdate>1974</pubdate>
  <volume>40</volume>
  <issue>6</issue>
  <fpage>747</fpage>
  <lpage>-755</lpage>
</bibl>

<bibl id="B9">
  <title><p>{The calcarine sulcus as an estimate of the total volume of human
  striate cortex: a morphometric study of reliability and intersubject
  variability}</p></title>
  <aug>
    <au><snm>Gilissen</snm><fnm>E</fnm></au>
    <au><snm>Zilles</snm><fnm>K</fnm></au>
  </aug>
  <source>J Hirnforsch</source>
  <pubdate>1996</pubdate>
  <volume>37</volume>
  <issue>1</issue>
  <fpage>57</fpage>
  <lpage>-66</lpage>
</bibl>

<bibl id="B10">
  <title><p>Correlated size variations in human visual cortex, lateral
  geniculate nucleus, and optic tract</p></title>
  <aug>
    <au><snm>Andrews</snm><fnm>T. J.</fnm></au>
    <au><snm>Halpern</snm><fnm>S. D.</fnm></au>
    <au><snm>Purves</snm><fnm>D.</fnm></au>
  </aug>
  <source>J Neurosci</source>
  <pubdate>1997</pubdate>
  <volume>17</volume>
  <fpage>2859</fpage>
  <lpage>2868</lpage>
</bibl>

<bibl id="B11">
  <title><p>{Area V5 of the human brain: evidence from a combined study using
  positron emission tomography and magnetic resonance imaging}</p></title>
  <aug>
    <au><snm>Watson</snm><fnm>J D</fnm></au>
    <au><snm>Myers</snm><fnm>R</fnm></au>
    <au><snm>Frackowiak</snm><fnm>R S</fnm></au>
    <au><snm>Hajnal</snm><fnm>J V</fnm></au>
    <au><snm>Woods</snm><fnm>R P</fnm></au>
    <au><snm>Mazziotta</snm><fnm>J C</fnm></au>
    <au><snm>Shipp</snm><fnm>S</fnm></au>
    <au><snm>Zeki</snm><fnm>S</fnm></au>
  </aug>
  <source>Cereb Cortex</source>
  <pubdate>1993</pubdate>
  <volume>3</volume>
  <issue>2</issue>
  <fpage>79</fpage>
  <lpage>-94</lpage>
</bibl>

<bibl id="B12">
  <title><p>{In vivo identification of human cortical areas using
  high-resolution MRI: an approach to cerebral structure-function
  correlation}</p></title>
  <aug>
    <au><snm>Walters</snm><fnm>NB</fnm></au>
    <au><snm>Egan</snm><fnm>GF</fnm></au>
    <au><snm>Kril</snm><fnm>JJ</fnm></au>
    <au><snm>Kean</snm><fnm>M</fnm></au>
    <au><snm>Waley</snm><fnm>P</fnm></au>
    <au><snm>Jenkinson</snm><fnm>M</fnm></au>
    <au><snm>Watson</snm><fnm>JDG</fnm></au>
  </aug>
  <source>Proc Natl Acad Sci U S A</source>
  <pubdate>2003</pubdate>
  <volume>100</volume>
  <issue>5</issue>
  <fpage>2981</fpage>
  <lpage>-2986</lpage>
</bibl>

<bibl id="B13">
  <title><p>{Functional topographic mapping of the cortical ribbon in human
  vision with conventional MRI scanners}</p></title>
  <aug>
    <au><snm>Schneider</snm><fnm>W</fnm></au>
    <au><snm>Noll</snm><fnm>D C</fnm></au>
    <au><snm>Cohen</snm><fnm>J D</fnm></au>
  </aug>
  <source>Nature</source>
  <pubdate>1993</pubdate>
  <volume>365</volume>
  <issue>6442</issue>
  <fpage>150</fpage>
  <lpage>-153</lpage>
</bibl>

<bibl id="B14">
  <title><p>{fMRI of human visual cortex}</p></title>
  <aug>
    <au><snm>Engel</snm><fnm>S. A.</fnm></au>
    <au><snm>Rumelhart</snm><fnm>D. E.</fnm></au>
    <au><snm>Wandell</snm><fnm>B. A.</fnm></au>
    <au><snm>Lee</snm><fnm>A. T.</fnm></au>
    <au><snm>Glover</snm><fnm>G. H.</fnm></au>
    <au><snm>Chichilnisky</snm><fnm>E. J.</fnm></au>
    <au><snm>Shadlen</snm><fnm>M. N.</fnm></au>
  </aug>
  <source>Nature</source>
  <pubdate>1994</pubdate>
  <volume>369</volume>
  <fpage>525</fpage>
</bibl>

<bibl id="B15">
  <title><p>{Borders of multiple visual areas in humans revealed by functional
  magnetic resonance imaging}</p></title>
  <aug>
    <au><snm>Sereno</snm><fnm>M. I.</fnm></au>
    <au><snm>Dale</snm><fnm>A. M.</fnm></au>
    <au><snm>Reppas</snm><fnm>J. B.</fnm></au>
    <au><snm>Kwong</snm><fnm>K. K.</fnm></au>
    <au><snm>Belliveau</snm><fnm>J. W.</fnm></au>
    <au><snm>Brady</snm><fnm>T. J.</fnm></au>
    <au><snm>Rosen</snm><fnm>B. R.</fnm></au>
    <au><snm>Tootell</snm><fnm>R. B.</fnm></au>
  </aug>
  <source>Science</source>
  <pubdate>1995</pubdate>
  <volume>268</volume>
  <fpage>889</fpage>
  <lpage>893</lpage>
</bibl>

<bibl id="B16">
  <title><p>{Mapping striate and extrastriate visual areas in human cerebral
  cortex}</p></title>
  <aug>
    <au><snm>DeYoe</snm><fnm>E A</fnm></au>
    <au><snm>Carman</snm><fnm>G J</fnm></au>
    <au><snm>Bandettini</snm><fnm>P</fnm></au>
    <au><snm>Glickman</snm><fnm>S</fnm></au>
    <au><snm>Wieser</snm><fnm>J</fnm></au>
    <au><snm>Cox</snm><fnm>R</fnm></au>
    <au><snm>Miller</snm><fnm>D</fnm></au>
    <au><snm>Neitz</snm><fnm>J</fnm></au>
  </aug>
  <source>Proc Natl Acad Sci U S A</source>
  <pubdate>1996</pubdate>
  <volume>93</volume>
  <issue>6</issue>
  <fpage>2382</fpage>
  <lpage>-2386</lpage>
</bibl>

<bibl id="B17">
  <title><p>{Retinotopic organization in human visual cortex and the spatial
  precision of functional MRI}</p></title>
  <aug>
    <au><snm>Engel</snm><fnm>S A</fnm></au>
    <au><snm>Glover</snm><fnm>G H</fnm></au>
    <au><snm>Wandell</snm><fnm>B A</fnm></au>
  </aug>
  <source>Cereb Cortex</source>
  <pubdate>1997</pubdate>
  <volume>7</volume>
  <issue>2</issue>
  <fpage>181</fpage>
  <lpage>-192</lpage>
</bibl>

<bibl id="B18">
  <title><p>{Estimating linear cortical magnification in human primary visual
  cortex via dynamic programming}</p></title>
  <aug>
    <au><snm>Qiu</snm><fnm>A</fnm></au>
    <au><snm>Rosenau</snm><fnm>BJ</fnm></au>
    <au><snm>Greenberg</snm><fnm>AS</fnm></au>
    <au><snm>Hurdal</snm><fnm>MK</fnm></au>
    <au><snm>Barta</snm><fnm>P</fnm></au>
    <au><snm>Yantis</snm><fnm>S</fnm></au>
    <au><snm>Miller</snm><fnm>MI</fnm></au>
  </aug>
  <source>Neuroimage</source>
  <pubdate>2006</pubdate>
  <volume>31</volume>
  <issue>1</issue>
  <fpage>125</fpage>
  <lpage>-138</lpage>
</bibl>

<bibl id="B19">
  <title><p>{{T}he topography of primate retina: a study of the human,
  bushbaby, and new- and old-world monkeys}</p></title>
  <aug>
    <au><snm>Stone</snm><fnm>J.</fnm></au>
    <au><snm>Johnston</snm><fnm>E.</fnm></au>
  </aug>
  <source>J. Comp. Neurol.</source>
  <pubdate>1981</pubdate>
  <volume>196</volume>
  <issue>2</issue>
  <fpage>205</fpage>
  <lpage>-223</lpage>
</bibl>

<bibl id="B20">
  <title><p>{Cortical magnification factor and the ganglion cell density of the
  primate retina}</p></title>
  <aug>
    <au><snm>W{\"a}ssle</snm><fnm>H</fnm></au>
    <au><snm>Gr{\"u}nert</snm><fnm>U</fnm></au>
    <au><snm>R{\"o}hrenbeck</snm><fnm>J</fnm></au>
    <au><snm>Boycott</snm><fnm>B B</fnm></au>
  </aug>
  <source>Nature</source>
  <pubdate>1989</pubdate>
  <volume>341</volume>
  <issue>6243</issue>
  <fpage>643</fpage>
  <lpage>-646</lpage>
</bibl>

<bibl id="B21">
  <title><p>{Topography of ganglion cells in human retina}</p></title>
  <aug>
    <au><snm>Curcio</snm><fnm>C A</fnm></au>
    <au><snm>Allen</snm><fnm>K A</fnm></au>
  </aug>
  <source>J Comp Neurol</source>
  <pubdate>1990</pubdate>
  <volume>300</volume>
  <issue>1</issue>
  <fpage>5</fpage>
  <lpage>-25</lpage>
</bibl>

<bibl id="B22">
  <title><p>{{I}sotropy of cortical magnification and topography of striate
  cortex}</p></title>
  <aug>
    <au><snm>Rovamo</snm><fnm>J.</fnm></au>
    <au><snm>Virsu</snm><fnm>V.</fnm></au>
  </aug>
  <source>Vision Res.</source>
  <pubdate>1984</pubdate>
  <volume>24</volume>
  <issue>3</issue>
  <fpage>283</fpage>
  <lpage>-286</lpage>
</bibl>

<bibl id="B23">
  <title><p>{{D}oes retinotopy influence cortical folding in primate visual
  cortex?}</p></title>
  <aug>
    <au><snm>Rajimehr</snm><fnm>R.</fnm></au>
    <au><snm>Tootell</snm><fnm>R. B.</fnm></au>
  </aug>
  <source>J. Neurosci.</source>
  <pubdate>2009</pubdate>
  <volume>29</volume>
  <issue>36</issue>
  <fpage>11149</fpage>
  <lpage>-11152</lpage>
</bibl>

<bibl id="B24">
  <title><p>{The distribution of rods and cones in the retina of the cat (Felis
  domesticus)}</p></title>
  <aug>
    <au><snm>Steinberg</snm><fnm>R H</fnm></au>
    <au><snm>Reid</snm><fnm>M</fnm></au>
    <au><snm>Lacy</snm><fnm>P L</fnm></au>
  </aug>
  <source>J Comp Neurol</source>
  <pubdate>1973</pubdate>
  <volume>148</volume>
  <issue>2</issue>
  <fpage>229</fpage>
  <lpage>-248</lpage>
</bibl>

<bibl id="B25">
  <title><p>{Contralateral visual field representation in area 17 of the
  cerebral cortex of the agouti: a comparison between the cortical
  magnification factor and retinal ganglion cell distribution}</p></title>
  <aug>
    <au><snm>Picanco Diniz</snm><fnm>C W</fnm></au>
    <au><snm>Silveira</snm><fnm>L C</fnm></au>
    <au><snm>Carvalho</snm><fnm>M S</fnm></au>
    <au><snm>Oswaldo Cruz</snm><fnm>E</fnm></au>
  </aug>
  <source>Neuroscience</source>
  <pubdate>1991</pubdate>
  <volume>44</volume>
  <issue>2</issue>
  <fpage>325</fpage>
  <lpage>-333</lpage>
</bibl>

<bibl id="B26">
  <title><p>{Topography of ganglion cells in the dog and wolf
  retina}</p></title>
  <aug>
    <au><snm>Peichl</snm><fnm>L</fnm></au>
  </aug>
  <source>J Comp Neurol</source>
  <pubdate>1992</pubdate>
  <volume>324</volume>
  <issue>4</issue>
  <fpage>603</fpage>
  <lpage>-620</lpage>
</bibl>

<bibl id="B27">
  <title><p>{Cytoarchitecture and visual field representation in area 17 of the
  tammar wallaby (Macropus eugenii)}</p></title>
  <aug>
    <au><snm>Vidyasagar</snm><fnm>T R</fnm></au>
    <au><snm>Wye Dvorak</snm><fnm>J</fnm></au>
    <au><snm>Henry</snm><fnm>G H</fnm></au>
    <au><snm>Mark</snm><fnm>R F</fnm></au>
  </aug>
  <source>J Comp Neurol</source>
  <pubdate>1992</pubdate>
  <volume>325</volume>
  <issue>2</issue>
  <fpage>291</fpage>
  <lpage>-300</lpage>
</bibl>

<bibl id="B28">
  <title><p>{Topography of ganglion cells in the retina of the
  horse}</p></title>
  <aug>
    <au><snm>Guo</snm><fnm>X</fnm></au>
    <au><snm>Sugita</snm><fnm>S</fnm></au>
  </aug>
  <source>J Vet Med Sci</source>
  <pubdate>2000</pubdate>
  <volume>62</volume>
  <issue>11</issue>
  <fpage>1145</fpage>
  <lpage>-1150</lpage>
</bibl>

<bibl id="B29">
  <title><p>{Topographies of retinal cone photoreceptors in two Australian
  marsupials}</p></title>
  <aug>
    <au><snm>Arrese</snm><fnm>C A</fnm></au>
    <au><snm>Rodger</snm><fnm>J</fnm></au>
    <au><snm>Beazley</snm><fnm>L D</fnm></au>
    <au><snm>Shand</snm><fnm>J</fnm></au>
  </aug>
  <source>Vis Neurosci</source>
  <pubdate>2003</pubdate>
  <volume>20</volume>
  <issue>3</issue>
  <fpage>307</fpage>
  <lpage>-311</lpage>
</bibl>

<bibl id="B30">
  <title><p>{Topography of photoreceptors and retinal ganglion cells in the
  spotted hyena (Crocuta crocuta)}</p></title>
  <aug>
    <au><snm>Calderone</snm><fnm>JB</fnm></au>
    <au><snm>Reese</snm><fnm>BE</fnm></au>
    <au><snm>Jacobs</snm><fnm>GH</fnm></au>
  </aug>
  <source>Brain Behav Evol</source>
  <pubdate>2003</pubdate>
  <volume>62</volume>
  <issue>4</issue>
  <fpage>182</fpage>
  <lpage>-192</lpage>
</bibl>

<bibl id="B31">
  <title><p>{Visual field anisotropy revealed by perceptual
  filling-in}</p></title>
  <aug>
    <au><snm>Sakaguchi</snm><fnm>Y</fnm></au>
  </aug>
  <source>Vision Res</source>
  <pubdate>2003</pubdate>
  <volume>43</volume>
  <issue>19</issue>
  <fpage>2029</fpage>
  <lpage>2038</lpage>
</bibl>

<bibl id="B32">
  <title><p>Patterns of Cerebral Integration inicated by Scotomas of
  Migraine</p></title>
  <aug>
    <au><snm>Lashley</snm><fnm>K</fnm></au>
  </aug>
  <source>Arch Neurol Psychiatry</source>
  <pubdate>1941</pubdate>
  <volume>46</volume>
  <fpage>331</fpage>
  <lpage>339</lpage>
</bibl>

<bibl id="B33">
  <title><p>Distortions of human retinotopy obtained with temporal phase mapped
  fMRI</p></title>
  <aug>
    <au><snm>Janik</snm><fnm>J.J.</fnm></au>
    <au><snm>Ropella</snm><fnm>K.M.</fnm></au>
    <au><snm>DeYoe</snm><fnm>E.A.</fnm></au>
  </aug>
  <source>Soc. Neurosci. Abstr</source>
  <pubdate>2003</pubdate>
  <volume>29</volume>
  <fpage>658.8</fpage>
</bibl>

<bibl id="B34">
  <title><p>{Deformation-based surface morphometry applied to gray matter
  deformation}</p></title>
  <aug>
    <au><snm>Chung</snm><fnm>MK</fnm></au>
    <au><snm>Worsley</snm><fnm>KJ</fnm></au>
    <au><snm>Robbins</snm><fnm>S</fnm></au>
    <au><snm>Paus</snm><fnm>T</fnm></au>
    <au><snm>Taylor</snm><fnm>J</fnm></au>
    <au><snm>Giedd</snm><fnm>JN</fnm></au>
    <au><snm>Rapoport</snm><fnm>JL</fnm></au>
    <au><snm>Evans</snm><fnm>AC</fnm></au>
  </aug>
  <source>Neuroimage</source>
  <pubdate>2003</pubdate>
  <volume>18</volume>
  <issue>2</issue>
  <fpage>198</fpage>
  <lpage>-213</lpage>
</bibl>

<bibl id="B35">
  <title><p>{{R}etinotopic maps, spatial tuning, and locations of human visual
  areas in surface coordinates characterized with multifocal and blocked
  {f}{M}{R}{I} designs}</p></title>
  <aug>
    <au><snm>Henriksson</snm><fnm>L.</fnm></au>
    <au><snm>Karvonen</snm><fnm>J.</fnm></au>
    <au><snm>Salminen Vaparanta</snm><fnm>N.</fnm></au>
    <au><snm>Railo</snm><fnm>H.</fnm></au>
    <au><snm>Vanni</snm><fnm>S.</fnm></au>
  </aug>
  <source>PLoS ONE</source>
  <pubdate>2012</pubdate>
  <volume>7</volume>
  <issue>5</issue>
  <fpage>e36859</fpage>
</bibl>

<bibl id="B36">
  <title><p>Handbook of Continuum Mechanics</p></title>
  <aug>
    <au><snm>Salencon</snm><fnm>J</fnm></au>
  </aug>
  <publisher>Springer, Berlin</publisher>
  <pubdate>2001</pubdate>
</bibl>

<bibl id="B37">
  <title><p>Simulation of Partial Visual Field Defects Using Selforganizing
  Maps of the Curved Surface of the Primary Visual Cortex</p></title>
  <aug>
    <au><snm>Tusch</snm><fnm>J</fnm></au>
  </aug>
  <source>Master's thesis</source>
  <publisher>Otto-von-Guericke University of Magdeburg</publisher>
  <pubdate>2004</pubdate>
</bibl>

<bibl id="B38">
  <title><p>{Overlap of receptive field centers and representation of the
  visual field in the cat's optic tract}</p></title>
  <aug>
    <au><snm>Fischer</snm><fnm>B</fnm></au>
  </aug>
  <source>Vision Res</source>
  <pubdate>1973</pubdate>
  <volume>13</volume>
  <issue>11</issue>
  <fpage>2113</fpage>
  <lpage>-2120</lpage>
</bibl>

<bibl id="B39">
  <title><p>The representation of the visual field in striate and adjoining
  cortex of the owl monkey ({A}otus trivirgatus)</p></title>
  <aug>
    <au><snm>Daniel</snm><fnm>P. M.</fnm></au>
    <au><snm>Whitteridge</snm><fnm>D.</fnm></au>
  </aug>
  <source>J. Physiol. (London)</source>
  <pubdate>1961</pubdate>
  <volume>159</volume>
  <fpage>203</fpage>
  <lpage>221</lpage>
</bibl>

<bibl id="B40">
  <title><p>Spatial mapping in primate sensory projection: analytic structure
  and relavance to perception</p></title>
  <aug>
    <au><snm>Schwartz</snm><fnm>E.</fnm></au>
  </aug>
  <source>Biol. Cybern.</source>
  <pubdate>1977</pubdate>
  <volume>25</volume>
  <fpage>181</fpage>
  <lpage>194</lpage>
</bibl>

<bibl id="B41">
  <title><p>Complex Analysis</p></title>
  <aug>
    <au><snm>Ahlfors</snm><fnm>LV</fnm></au>
  </aug>
  <publisher>McGraw-Hill Book Company</publisher>
  <pubdate>1953</pubdate>
</bibl>

<bibl id="B42">
  <title><p>{Electrophysiological estimate of human cortical
  magnification}</p></title>
  <aug>
    <au><snm>Slotnick</snm><fnm>S. D.</fnm></au>
    <au><snm>Klein</snm><fnm>S. A.</fnm></au>
    <au><snm>Carney</snm><fnm>T.</fnm></au>
    <au><snm>Sutter</snm><fnm>E. E.</fnm></au>
  </aug>
  <source>Clin Neurophysiol</source>
  <pubdate>2001</pubdate>
  <volume>112</volume>
  <issue>7</issue>
  <fpage>1349</fpage>
  <lpage>1356</lpage>
  <note>Clinical Trial</note>
</bibl>

<bibl id="B43">
  <title><p>Human magnification factor and its relationto visual
  acuity</p></title>
  <aug>
    <au><snm>Cowey</snm><fnm>A.</fnm></au>
    <au><snm>Rolls</snm><fnm>E. T.</fnm></au>
  </aug>
  <source>Exp. Brain Res.</source>
  <pubdate>1974</pubdate>
  <volume>3</volume>
  <fpage>447</fpage>
  <lpage>454</lpage>
</bibl>

<bibl id="B44">
  <title><p>A Geometric Method for Automatic Extraction of Sulcal
  Fundi</p></title>
  <aug>
    <au><snm>Kao</snm><fnm>C.</fnm></au>
    <au><snm>Hofer</snm><fnm>M.</fnm></au>
    <au><snm>Sapiro</snm><fnm>G.</fnm></au>
    <au><snm>Stern</snm><fnm>J.</fnm></au>
    <au><snm>Rottenberg</snm><fnm>D. A.</fnm></au>
  </aug>
  <source>Proc. ISBI 06, IEEE</source>
  <pubdate>2006</pubdate>
  <fpage>1168</fpage>
  <lpage>1171</lpage>
</bibl>

<bibl id="B45">
  <title><p>{Migraine aura dynamics after reverse retinotopic mapping of weak
  excitation waves in the primary visual cortex}</p></title>
  <aug>
    <au><snm>Dahlem</snm><fnm>M. A.</fnm></au>
    <au><snm>M{\"u}ller</snm><fnm>S. C.</fnm></au>
  </aug>
  <source>Biol Cybern</source>
  <pubdate>2003</pubdate>
  <volume>88</volume>
  <fpage>419</fpage>
  <lpage>424</lpage>
</bibl>

<bibl id="B46">
  <title><p>Migraine aura: retracting particle-like waves in weakly susceptible
  cortex</p></title>
  <aug>
    <au><snm>Dahlem</snm><fnm>M. A.</fnm></au>
    <au><snm>Hadjikhani</snm><fnm>N.</fnm></au>
  </aug>
  <source>PLoS ONE</source>
  <pubdate>2009</pubdate>
  <volume>4</volume>
  <fpage>e5007</fpage>
</bibl>

<bibl id="B47">
  <title><p>{Migraine phosphenes and the retino-cortical magnification
  factor}</p></title>
  <aug>
    <au><snm>Gr{\"u}sser</snm><fnm>O. J.</fnm></au>
  </aug>
  <source>Vision Res</source>
  <pubdate>1995</pubdate>
  <volume>35</volume>
  <fpage>1125</fpage>
  <lpage>1134</lpage>
</bibl>

<bibl id="B48">
  <title><p>{Reorganization of human cortical maps caused by inherited
  photoreceptor abnormalities}</p></title>
  <aug>
    <au><snm>Baseler</snm><fnm>HA</fnm></au>
    <au><snm>Brewer</snm><fnm>AA</fnm></au>
    <au><snm>Sharpe</snm><fnm>LT</fnm></au>
    <au><snm>Morland</snm><fnm>AB</fnm></au>
    <au><snm>Jagle</snm><fnm>H</fnm></au>
    <au><snm>Wandell</snm><fnm>BA</fnm></au>
  </aug>
  <source>Nat Neurosci</source>
  <pubdate>2002</pubdate>
  <volume>5</volume>
  <issue>4</issue>
  <fpage>364</fpage>
  <lpage>-370</lpage>
</bibl>

<bibl id="B49">
  <title><p>{Symmetry of cortical folding abnormalities in Williams syndrome
  revealed by surface-based analyses}</p></title>
  <aug>
    <au><snm>Van Essen</snm><fnm>DC</fnm></au>
    <au><snm>Dierker</snm><fnm>D</fnm></au>
    <au><snm>Snyder</snm><fnm>A Z</fnm></au>
    <au><snm>Raichle</snm><fnm>ME</fnm></au>
    <au><snm>Reiss</snm><fnm>AL</fnm></au>
    <au><snm>Korenberg</snm><fnm>J</fnm></au>
  </aug>
  <source>J Neurosci</source>
  <pubdate>2006</pubdate>
  <volume>26</volume>
  <fpage>5470</fpage>
  <lpage>-5483</lpage>
</bibl>

<bibl id="B50">
  <title><p>{{C}ortical magnification within human primary visual cortex
  correlates with acuity thresholds}</p></title>
  <aug>
    <au><snm>Duncan</snm><fnm>R.O.</fnm></au>
    <au><snm>Boynton</snm><fnm>G.M.</fnm></au>
  </aug>
  <source>Neuron</source>
  <pubdate>2003</pubdate>
  <volume>38</volume>
  <fpage>659</fpage>
  <lpage>-671</lpage>
</bibl>

<bibl id="B51">
  <title><p>Co-planar Stereotaxic Atlas of the Human Brain</p></title>
  <aug>
    <au><snm>Talairach</snm><fnm>J.</fnm></au>
    <au><snm>Tournoux</snm><fnm>P.</fnm></au>
  </aug>
  <publisher>Thieme Medical, New York</publisher>
  <pubdate>1988</pubdate>
</bibl>

<bibl id="B52">
  <title><p>{The fusiform face area: a module in human extrastriate cortex
  specialized for face perception}</p></title>
  <aug>
    <au><snm>Kanwisher</snm><fnm>N</fnm></au>
    <au><snm>McDermott</snm><fnm>J</fnm></au>
    <au><snm>Chun</snm><fnm>M M</fnm></au>
  </aug>
  <source>J Neurosci</source>
  <pubdate>1997</pubdate>
  <volume>17</volume>
  <issue>11</issue>
  <fpage>4302</fpage>
  <lpage>-4311</lpage>
</bibl>

<bibl id="B53">
  <title><p>{A cortical representation of the local visual
  environment}</p></title>
  <aug>
    <au><snm>Epstein</snm><fnm>R</fnm></au>
    <au><snm>Kanwisher</snm><fnm>N</fnm></au>
  </aug>
  <source>Nature</source>
  <pubdate>1998</pubdate>
  <volume>392</volume>
  <issue>6676</issue>
  <fpage>598</fpage>
  <lpage>-601</lpage>
</bibl>

<bibl id="B54">
  <title><p>{Domain Specificity in Visual Cortex}</p></title>
  <aug>
    <au><snm>Downing</snm><fnm>PE</fnm></au>
    <au><snm>Chan</snm><fnm>AW</fnm></au>
    <au><snm>Peelen</snm><fnm>MV</fnm></au>
    <au><snm>Dodds</snm><fnm>CM</fnm></au>
    <au><snm>Kanwisher</snm><fnm>N</fnm></au>
  </aug>
  <source>Cereb Cortex</source>
  <pubdate>2005</pubdate>
  <volume>EPrint</volume>
  <note>JOURNAL ARTICLE</note>
</bibl>

<bibl id="B55">
  <title><p>The {V}1-{V}2-{V}3 complex: quasiconformal dipole maps in primate
  striate and extra-striate cortex</p></title>
  <aug>
    <au><snm>Balasubramanian</snm><fnm>M.</fnm></au>
    <au><snm>Polimeni</snm><fnm>J.</fnm></au>
    <au><snm>Schwartz</snm><fnm>E.L.</fnm></au>
  </aug>
  <source>Neural Netw.</source>
  <pubdate>2002</pubdate>
  <volume>15</volume>
  <fpage>1157</fpage>
  <lpage>1163</lpage>
</bibl>

<bibl id="B56">
  <title><p>{Deoxyglucose analysis of retinotopic organization in primate
  striate cortex}</p></title>
  <aug>
    <au><snm>Tootell</snm><fnm>R B</fnm></au>
    <au><snm>Silverman</snm><fnm>M S</fnm></au>
    <au><snm>Switkes</snm><fnm>E</fnm></au>
    <au><snm>De Valois</snm><fnm>R L</fnm></au>
  </aug>
  <source>Science</source>
  <pubdate>1982</pubdate>
  <volume>218</volume>
  <issue>4575</issue>
  <fpage>902</fpage>
  <lpage>-904</lpage>
</bibl>

<bibl id="B57">
  <title><p>{Why the cortical magnification factor in rhesus can not be
  isotropic}</p></title>
  <aug>
    <au><snm>Sakitt</snm><fnm>B</fnm></au>
  </aug>
  <source>Vision Res</source>
  <pubdate>1982</pubdate>
  <volume>22</volume>
  <issue>3</issue>
  <fpage>417</fpage>
  <lpage>-421</lpage>
</bibl>

<bibl id="B58">
  <title><p>Mapping of cytochrome oxidase patches and ocular dominance columns
  in human visual cortex.</p></title>
  <aug>
    <au><snm>Horton</snm><fnm>J. C.</fnm></au>
    <au><snm>Hedley Whyte</snm><fnm>E. T.</fnm></au>
  </aug>
  <source>Philos. Trans. R. Soc. Lond. B Biol. Sci.</source>
  <pubdate>1984</pubdate>
  <volume>304</volume>
  <fpage>255</fpage>
  <lpage>72</lpage>
</bibl>

<bibl id="B59">
  <title><p>{The complete pattern of ocular dominance stripes in the striate
  cortex and visual field of the macaque monkey}</p></title>
  <aug>
    <au><snm>LeVay</snm><fnm>S</fnm></au>
    <au><snm>Connolly</snm><fnm>M</fnm></au>
    <au><snm>Houde</snm><fnm>J</fnm></au>
    <au><snm>Van Essen</snm><fnm>D C</fnm></au>
  </aug>
  <source>J Neurosci</source>
  <pubdate>1985</pubdate>
  <volume>5</volume>
  <issue>2</issue>
  <fpage>486</fpage>
  <lpage>-501</lpage>
</bibl>

<bibl id="B60">
  <title><p>External Morphology of the Primate Brain</p></title>
  <aug>
    <au><snm>Connolly</snm><fnm>C. J.</fnm></au>
  </aug>
  <publisher>Springfield, Ill., C.C. Thomas</publisher>
  <pubdate>1950</pubdate>
</bibl>

<bibl id="B61">
  <title><p>Retinotopic organization of human visual cortex: departures from
  the classical model</p></title>
  <aug>
    <au><snm>Aine</snm><fnm>C. J.</fnm></au>
    <au><snm>Supek</snm><fnm>S.</fnm></au>
    <au><snm>George</snm><fnm>J. S.</fnm></au>
    <au><snm>Ranken</snm><fnm>D.</fnm></au>
    <au><snm>Lewine</snm><fnm>J.</fnm></au>
    <au><snm>Sanders</snm><fnm>J.</fnm></au>
    <au><snm>Best</snm><fnm>E.</fnm></au>
    <au><snm>Tiee</snm><fnm>W.</fnm></au>
    <au><snm>Flynn</snm><fnm>E. R.</fnm></au>
    <au><snm>Wood</snm><fnm>C. C.</fnm></au>
  </aug>
  <source>Cereb. Cortex</source>
  <pubdate>1996</pubdate>
  <volume>6</volume>
  <fpage>354</fpage>
  <lpage>361</lpage>
</bibl>

<bibl id="B62">
  <title><p>{Olfactory receptors and odor coding in mammals}</p></title>
  <aug>
    <au><snm>Buck</snm><fnm>LB</fnm></au>
  </aug>
  <source>Nutr Rev</source>
  <pubdate>2004</pubdate>
  <volume>62</volume>
  <issue>11 Pt 2</issue>
  <fpage>184</fpage>
  <lpage>-188</lpage>
</bibl>

<bibl id="B63">
  <title><p>Theoretical Neuroscience: Computational and Mathematical Modeling
  of Neural Systems</p></title>
  <aug>
    <au><snm>Dayan</snm><fnm>P</fnm></au>
    <au><snm>Abbott</snm><fnm>L. F.</fnm></au>
  </aug>
  <publisher>{The MIT Press}</publisher>
  <pubdate>2001</pubdate>
</bibl>

</refgrp>
} 
}     


\ifthenelse{\boolean{publ}}{\end{multicols}}{}



%
%

%
%




\clearpage
\pagebreak

\end{bmcformat}
\end{document}